\documentclass[prb,twocolumn,showpacs,superscriptaddress]{revtex4}
\usepackage{amsmath,amssymb,graphicx,color,bm,soul}

\begin{document}

\title{Ballistic spin transport in exciton gases}
\author{A. V. Kavokin}
\affiliation{Laboratoire Charles Coulomb, CNRS-Universite de Montpellier II, Pl. Eugene
de Bataillon, 34095 Montpellier Cedex, France}
\affiliation{Physics and Astronomy School, University of Southampton, Highfield,
Southampton, SO171BJ, UK}
\author{M. Vladimirova}
\affiliation{Laboratoire Charles Coulomb, CNRS-Universite de Montpellier II, Pl. Eugene
de Bataillon, 34095 Montpellier Cedex, France}
\author{B. Jouault}
\affiliation{Laboratoire Charles Coulomb, CNRS-Universite de Montpellier II, Pl. Eugene
de Bataillon, 34095 Montpellier Cedex, France}
\author{T. C. H. Liew}
\affiliation{Mediterranean Institute of Fundamental Physics, 31, via Appia Nuova, Roma,
00040, Italy}
\author{J.R.~Leonard}
\affiliation{Department of Physics, University of California at San Diego, La Jolla, CA
92093-0319, USA}
\author{L.V.~Butov}
\affiliation{Department of Physics, University of California at San Diego, La Jolla, CA
92093-0319, USA}

\begin{abstract}
Traditional spintronics relies on spin transport by charge carriers, such as
electrons in semiconductor crystals. The challenges for the realization of
long-range electron spin transport include rapid spin relaxation due to
electron scattering. Scattering and, in turn, spin relaxation can be
effectively suppressed in excitonic devices where the spin currents are
carried by electrically neutral bosonic quasi-particles: excitons or
exciton-polaritons. They can form coherent quantum liquids that carry spins
over macroscopic distances. The price to pay is a finite life-time of the
bosonic spin carriers. We present the theory of exciton ballistic spin
transport which may be applied to a range of systems supporting bosonic spin
transport, in particular, to indirect excitons in coupled quantum wells. We
describe the effect of spin-orbit interaction for the electron and the hole
on the exciton spin, account for the Zeeman effect induced by external
magnetic fields and long range and short range exchange splittings of the
exciton resonances. We also consider exciton transport in the non-linear
regime and discuss the definitions of the exciton spin current, polarization
current and spin conductivity.
\end{abstract}

\maketitle



\section{Introduction}

Excitons are electrically neutral and have finite lifetimes. These are two
obstacles which make the development of excitonic spintronics, or
spin-optronics challenging. How possibly one can explore the current, which
is carried by neutral particles, and whose amplitude changes with distance
and time? - is a fair question to ask. While electrons and holes have been
considered as perfectly valid spin carriers, and exotic effects like the
spin Hall effect~\cite{Dyakonov1971} have been intensively studied for them~%
\cite{Sinova2004,Mishchenko2004,Kato2004,Wunderlich2005,Valenzuela2006}, the
spin currents carried by excitons~\cite{Leonard2009, High2012, High2013} and
exciton-polaritons~\cite{Leyder2007,Kammann2012} over tens or even hundreds
of micrometers remained relatively less explored. There existed a huge
imbalance of theoretical works on fermionic and bosonic spin transport. This
is changing now. A number of phenomena have been observed and studied in the
field of bosonic spin currents recently~\cite{Leonard2009}$^{-}$\cite%
{Wertz2012}. To summarize tens of publications in one sentence: bosonic
systems bring new quantum coherent effects to the physics of spin transport.
For instance, stimulation~\cite{Szymanska2010,Amo2009,Adrados2011} and
amplification~\cite{Wertz2012} of spin currents
are possible in exciton and exciton-polariton Bose gases. Bosonic
spintronics or spin-optronics operates with electrically neutral
spin carriers which makes control of spin currents carried by
excitons a non-trivial task. Fortunately, the exciton density
replaces charge in many aspects: the density currents may be
efficiently controlled by stationary or
dynamic potential gradients as demonstrated in recent works \cite%
{Kuznetsova2010, Leonard2012}. Combined with evident advantages of bosonic
amplification and low dephasing, this makes spin-optronics a valuable
alternative to fermionic spintronics. Besides bosonic effects, exciton spin
transport has another important specific feature: it is dissipative by its
nature, as the spin carriers have a finite (and short for excitons in
regular materials) life-time. In continuous wave optical experiments
stationary spin textures can appear: excitons are injected in the structure,
they propagate ballistically or diffusively, and eventually disappear by
radiative recombination. Their polarization properties and spin are
inherited by the emitted photons, that is why the polarization patterns
observed in near field photoluminescence experiments directly characterize
exciton spin currents in the plane of the structure.

The goal of this work is to define what the exciton spin, magnetization and
polarization currents are, and to explain how they can be described within
the most frequently used spin density matrix (DM) approach and mean-field~
\cite{Carusotto2004,Shelykh2006} approximation. We consider a specific
system, namely a planar zinc-blend semiconductor structure containing
quantum wells, where excitons can be formed. This choice is motivated by
recent experimental results in GaAs/AlGaAs based coupled quantum wells. We
limit the scope of this paper to heavy-hole excitons, however, our approach
can be easily extended to light-hole excitons or excitons in quantum wells
of a different symmetry. We do not speak here about the large variety of
recent experimental results and application of the formalism presented here
to the description of one particular experiment, as this would make this
paper too long and too specific. For a direct comparison of theoretical
simulations with the experimental data we address the reader to Ref. %
\onlinecite{High2013}. The approaches formulated here are suitable for the
description of a variety of excitonic spin effects in quantum wells.

The paper is organized as follows. In Section II we introduce the spin DM
formalism accounting for the different mechanisms of spin re-orientation and
the relation to electron and hole spin currents. In Section III we present
numerical results obtained within the spin DM formalism and analyze them. In
Section \ref{sec:GP} we study the non-linear spin dynamics of propagating
excitons using the Gross-Pitaevskii (GP) equations. The next three sections
of the paper are devoted to exciton spin currents and polarization currents.
Conclusions are perspectives are given in Section VIII.

\section{The Spin Matrix Formalism for Propagating Excitons}

In zinc-blend semiconductor quantum wells (e.g. in the most popular
GaAs/AlGaAs system), the lowest energy exciton states are formed by
electrons with spin projections on the structure axis of +1/2 and -1/2 and
heavy holes whose quasi-spin (sum of spin and orbital momentum) projection
to the structure axis is +3/2 or -3/2. Consequently, the exciton spin
defined as the sum of the electron spin and heavy-hole quasi-spin may have
one of four projections on the structure axis:
+1,-1,+2,-2
~\cite {Ivchenko1997}. These states are usually nearly degenerate,
while there may be some splitting between them due to the short and
long-range exchange interactions. Only the states with quasi-spin
projections $\pm$1 are coupled to the light, these are so-called
bright states. The states with quasi-spin projections $\pm$2 are
called dark states.

It is important to note that the present formalism addresses the spin part
of the exciton wavefunction, which is a product of electron and hole spin
functions. For example, the probability to find the exciton in the spin
state +1 is given by a product of probabilities to find an electron in the
spin state -1/2 and the heavy hole in the spin state +3/2. The four
component exciton wave-function is:
\begin{align}
\Psi & =\left( \Psi _{+1},\Psi _{-1},\Psi _{+2},\Psi _{-2}\right)  \notag \\
& =\left( \Psi _{e,-\frac{1}{2}}\Psi _{h,+\frac{3}{2}},\Psi _{e,+\frac{1}{2}%
}\Psi _{h,-\frac{3}{2}},\right.  \notag \\
&\hspace{30mm}\left.\Psi _{e,+\frac{1}{2}}\Psi _{h,+\frac{3}{2}},\Psi _{e,-%
\frac{1}{2}}\Psi _{h,-\frac{3}{2}}\right)
\end{align}%
where $\Psi _{e,+\frac{1}{2}}$ and $\Psi _{e,-\frac{1}{2}}$ are the
components of the electron spinor wavefunction; $\Psi _{h,+\frac{3}{2}}$ and
$\Psi _{h,-\frac{3}{2}}$ are the components of the heavy-hole spinor
wavefunction.

To describe the dynamics of the system we will first define the Hamiltonian
(Section~\ref{sec:Hamiltonian}) describing the different physical mechanisms
of spin evolution. We then introduce the spin DM (Section~\ref%
{sec:DensityMatrix}) for the description of exciton spin states. We relate
its components to the observable Stokes' vectors of light emitted by bright
excitons and use the Liouville equation to describe its evolution in time.
From the DM one can describe the polarization state of excitons and thus
exciton spin currents. It is also instructive to consider the relationship
with electron and hole spin currents (Section~\ref{sec:ehcurrents}).

\subsection{Hamiltonian}

\label{sec:Hamiltonian} Here we derive the exciton Hamiltonian in
the basis of +1,-1,+2,-2 states, accounting for the spin-orbit
interaction (Dresselhaus and Rashba
effects)~\cite{Rashba1988,Luo2010}, long- and short-range exchange
interactions~\cite{Ivchenko2005} and Zeeman effect, but neglecting
exciton-exciton interactions, which will be discussed in the
Section~\ref{sec:GP} and neglecting magnetic field effect on
center-of-mass
motion and internal structure of exciton \cite{Gorkov68,Lerner80,Lozovik2002}%
. We consider excitons propagating ballistically in the plane of a quantum
well. We shall characterize them by a fixed wave-vector, $\mathbf{k}_{%
\mathrm{ex}}$. We represent the full exciton Hamiltonian $\hat{\mathcal{H}}%
_{ex}^{tot}$ as a sum of three parts describing the spin-orbit and Zeeman
effects on electrons, $\hat{\mathcal{H}}_{ex}^{e}$, and holes, $\hat{%
\mathcal{H}}_{ex}^{h}$, and the exchange induced splittings of exciton
states, $\hat{\mathcal{H}}_{ex}^{ex}$:
\begin{equation}
\hat{\mathcal{H}}_{ex}^{tot}=\hat{\mathcal{H}}_{ex}^{e}+\hat{\mathcal{H}}%
_{ex}^{h}+\hat{\mathcal{H}}_{ex}^{ex}.
\end{equation}

\subsubsection{Dresselhaus and Zeeman terms}

We recall that the Rashba-Dresselhaus effect is a momentum-dependent
splitting of spin bands in two-dimensional semiconductor systems. It
originates from a combined effect of the atomic spin orbit coupling
and asymmetry of the potential in the direction perpendicular to the
two-dimensional plane. This asymmetry comes either from the applied
bias (which is described by the Rashba term in the Hamiltonian) or
from the intrinsic asymmetry of the crystal lattice (described by
the Dresselhaus term in the Hamiltonian). We shall separately
consider both the Dresselhaus term (in this sub-section) and the
Rashba term (in the next sub-section).

In order to build the $4\times4$ matrix Hamiltonian for excitons, we start
with simpler $2\times2$ Hamiltonians describing the spin-orbit and Zeeman
effects for electrons and holes.

The electron Hamiltonian in the basis of (+1/2,-1/2) spin states is:
\begin{equation}
\hat{\mathcal{H}}_{e}=\beta _{e}\left( k_{e,x}\hat{\sigma}_{x}-k_{e,y}\hat{%
\sigma}_{y}\right) -\frac{1}{2}g_{e}\mu _{B}\mathbf{B\hat{\sigma}}.
\label{eq:He1}
\end{equation}%
Here $g_{e}$ is the electron g-factor, $\mu _{B}$ is the Bohr magneton, $%
\mathbf{B}$ is a magnetic field, $\mathbf{\hat{\sigma}}$ is the Pauli matrix
vector, and $\beta _{e}$ is the Dresselhaus constant describing spin-orbit
interactions of electrons. The Pauli matrix operators are:
\begin{equation}
\hat{\sigma}_{z}=\left[
\begin{array}{cc}
1 & 0 \\
0 & -1%
\end{array}%
\right] ,\hspace{5mm}\hat{\sigma}_{y}=\left[
\begin{array}{cc}
0 & -i \\
i & 0%
\end{array}%
\right] ,\hspace{5mm}\hat{\sigma}_{x}=\left[
\begin{array}{cc}
0 & 1 \\
1 & 0%
\end{array}%
\right] .
\end{equation}%
Rewriting Eq.~(\ref{eq:He1}) and retaining only $z$-component of the
magnetic field, which corresponds to the Faraday geometry, one can obtain:
\begin{align}
\hat{\mathcal{H}}_{e}& =\left[
\begin{array}{cc}
-\frac{1}{2}g_{e}\mu _{B}B & \beta _{e}\left( k_{e,x}+ik_{e,y}\right) \\
\beta _{e}\left( k_{e,x}-ik_{e,y}\right) & \frac{1}{2}g_{e}\mu _{B}B%
\end{array}%
\right]  \notag \\
& =\left[
\begin{array}{cc}
-\frac{1}{2}g_{e}\mu _{B}B & \beta _{e}k_{e}e^{i\phi } \\
\beta _{e}k_{e}e^{-i\phi } & \frac{1}{2}g_{e}\mu _{B}B%
\end{array}%
\right] ,  \label{eq:He}
\end{align}%
where $\phi $ is the angle between $\mathbf{k}_{\mathrm{ex}}$ and the chosen
x-axis. The exciton Hamiltonian needs to be written in the basis of
(+1,-1,+2,-2) exciton spin states, which correspond to (-1/2, +1/2, +1/2,
-1/2) electron spin states. The electron spin-flip couples +1 and +2 states
as well as -1 and -2 states. For each of these two couples of states we
apply the Hamiltonian [Eq.~(\ref{eq:He})], which results in the following
electronic contribution to the $4\times 4$ exciton Hamiltonian:
\begin{equation}
\hat{\mathcal{H}}_{ex}^{e}=\left[
\begin{array}{cccc}
\frac{1}{2}g_{e}\mu _{B}B & 0 & \beta _{e}k_{e}e^{-i\phi } & 0 \\
0 & -\frac{1}{2}g_{e}\mu _{B}B & 0 & \beta _{e}k_{e}e^{i\phi } \\
\beta _{e}k_{e}e^{i\phi } & 0 & -\frac{1}{2}g_{e}\mu _{B}B & 0 \\
0 & \beta _{e}k_{e}e^{-i\phi } & 0 & \frac{1}{2}g_{e}\mu _{B}B%
\end{array}%
\right]  \label{eq:He4x4}
\end{equation}

\subsubsection{Rashba terms}

We note that another possible spin-orbit contribution to the Hamiltonian may
come from the Rashba effect, which takes place in biased quantum wells. The
Rashba term to be added in Eq.~(\ref{eq:He1}) is $\alpha _{e}\left( \hat{%
\sigma}_{x}k_{e,y}-\hat{\sigma}_{y}k_{e,x}\right) $, where $\alpha _{e}$ is
a constant proportional to the Rashba field. The contribution of the Rashba
term to the electron Hamiltonian, in the basis of (+1/2,-1/2) electron spin
states, can be re-written:
\begin{equation}
\hat{\mathcal{H}}_{e}^{\prime }=\left[
\begin{array}{cc}
0 & i\alpha _{e}k_{e}e^{-i\phi } \\
-i\alpha _{e}k_{e}e^{i\phi } & 0%
\end{array}%
\right] .  \label{eq:HeRashba1}
\end{equation}%
Using the same procedure as for Dresselhaus terms, this gives an additional
contribution to the exciton Hamiltonian, in the basis of (-1/2, +1/2, +1/2,
-1/2) electron spin states:
\begin{equation}
\hat{\mathcal{H}}_{ex}^{e\prime }=\left[
\begin{array}{cccc}
0 & 0 & -i\alpha _{e}k_{e}e^{i\phi } & 0 \\
0 & 0 & 0 & i\alpha _{e}k_{e}e^{-i\phi } \\
i\alpha _{e}k_{e}e^{-i\phi } & 0 & 0 & 0 \\
0 & -i\alpha _{e}k_{e}e^{i\phi } & 0 & 0%
\end{array}%
\right]
\end{equation}%
Unless stated explicitly, we will for simplicity omit the Rashba terms in
the rest of this paper and consider only the Dresselhaus terms. Note that
there are no linear in wave-vector Rashba terms for heavy holes in
zinc-blend quantum wells grown along the (001)-axis.

\subsubsection{Heavy-hole contribution (Faraday geometry)}

The heavy hole contribution to the Hamiltonian can be calculated with the
reasoning similar to the electron case. The heavy-hole Hamiltonian written
in the basis of (+3/2,-3/2) states is:
\begin{equation}
\hat{\mathcal{H}}_{h}=\beta _{h}\left( k_{h,x}\hat{\sigma}_{x}+k_{h,y}\hat{%
\sigma}_{y}\right) -\frac{1}{2}g_{h}\mu _{B}B\hat{\sigma}_{z}.
\label{eq:Hh1}
\end{equation}%
Here $g_{h}$ is the heavy-hole g-factor and $\beta _{h}$ is the
Dresselhaus constant for heavy holes~\cite{Rashba1988,Luo2010}. Note
that the Dresselhaus Hamiltonian is different for heavy holes formed
by $p$-orbital states and for conduction band electrons formed by
\textit{s-}orbital electronic states in a zinc-blend crystal
lattice. The resulting from Dresselhaus coupling effective magnetic
fields acting upon electron and
heavy hole spins are oriented differently as well. Re-writing Eq.~(\ref%
{eq:Hh1}), we obtain:
\begin{align}
\hat{\mathcal{H}}_{h}& =\left[
\begin{array}{cc}
-\frac{1}{2}g_{h}\mu _{B}B & \beta _{h}\left( k_{h,x}-ik_{h,y}\right) \\
\beta _{h}\left( k_{h,x}+ik_{h,y}\right) & \frac{1}{2}g_{h}\mu _{B}B%
\end{array}%
\right]  \notag \\
& =\left[
\begin{array}{cc}
-\frac{1}{2}g_{h}\mu _{B}B & \beta _{h}k_{h}e^{-i\phi } \\
\beta _{h}k_{h}e^{i\phi } & \frac{1}{2}g_{h}\mu _{B}B%
\end{array}%
\right] ,  \label{eq:Hh}
\end{align}%
The hole spin-flip couples +1 and -2 states as well as -1 and +2 states. For
each of these two couples of states we apply the Hamiltonian (\ref{eq:Hh}),
which results in the following hole contribution to the $4\times 4$ exciton
Hamiltonian:
\begin{equation}
\hat{\mathcal{H}}_{ex}^{h}=\left[
\begin{array}{cccc}
-\frac{1}{2}g_{h}\mu _{B}B & 0 & 0 & \beta _{h}k_{h}e^{-i\phi } \\
0 & \frac{1}{2}g_{h}\mu _{B}B & \beta _{h}k_{h}e^{i\phi } & 0 \\
0 & \beta _{h}k_{h}e^{-i\phi } & -\frac{1}{2}g_{h}\mu _{B}B & 0 \\
\beta _{h}k_{h}e^{i\phi } & 0 & 0 & \frac{1}{2}g_{h}\mu _{B}B%
\end{array}%
\right]  \label{eq:Hh4x4}
\end{equation}

\subsubsection{In-plane magnetic field (Voight geometry)}

If the magnetic field is applied in the plane, it splits electron and hole
states polarized in the plane of the quantum wells. Suppose that the field
is applied in the x-direction. In the electron and hole basis the Zeeman
Hamiltonian is in this case:
\begin{equation}
\hat{\mathcal{H}}_{e,h}=-\frac{1}{2}g_{e,h}\mu _{B}B\hat{\sigma}_{x}=\left[
\begin{array}{cc}
0 & -\frac{1}{2}g_{e,h}\mu _{B}B \\
-\frac{1}{2}g_{e,h}\mu _{B}B & 0%
\end{array}%
\right] .
\end{equation}%
Note that the hole $g$-factor in plane of the quantum well is different from
the $g$-factor in Faraday configuration, in general. This maps into the $%
\left( +1,-1,+2,-2\right) $ exciton basis as a Zeeman Hamiltonian of the
form:
\begin{equation}
\hat{\mathcal{H}}_{Z}=-\frac{\mu _{B}B}{2}\left[
\begin{array}{cccc}
0 & 0 & g_{e} & g_{h} \\
0 & 0 & g_{h} & g_{e} \\
g_{e} & g_{h} & 0 & 0 \\
g_{h} & g_{e} & 0 & 0 \\
&  &  &
\end{array}%
\right]
\end{equation}

\subsubsection{Exchange terms}

Besides the contributions from electron and hole spin orbit interactions and
Zeeman splitting, there may be a purely excitonic contribution to the
Hamiltonian, which is composed from the Hamiltonian for bright excitons
written in the basis (+1,-1):
\begin{equation}
\hat{\mathcal{H}}_{b}=E_{b}\hat{I}-\delta _{b}\hat{\sigma}_{x}=\left[
\begin{array}{cc}
E_{b} & -\delta _{b} \\
-\delta _{b} & E_{b}%
\end{array}%
\right]  \label{eq:Hb}
\end{equation}%
and the Hamiltonian for dark excitons written in the basis (+2,-2):
\begin{equation}
\hat{\mathcal{H}}_{d}=E_{d}\hat{I}-\delta _{d}\hat{\sigma}_{x}=\left[
\begin{array}{cc}
E_{d} & -\delta _{d} \\
-\delta _{d} & E_{d}%
\end{array}%
\right] ,  \label{eq:Hd}
\end{equation}%
where $\hat{I}$ is the identity matrix. The terms with $\delta _{b}$ and $%
\delta _{d}$ describe the splittings of bright and dark states
polarized along x and y axes in the plane of the structure due to
the long-range exchange interaction. The structural anisotropy is
virtually inevitable even in the best quality epitaxially grown
quantum wells. It arises from the reduced symmetry of
heterointerfaces, from local strains and from islands of quantum
well with fluctuations elongated in certain crystallographic
directions. $E_{b}-E_{d}$ is the splitting between bright (+1 and
-1) and dark (+2 and -2) exciton states due to the short-range
exchange interaction. In microcavities, this splitting is
additionally enhanced due to the vacuum field Rabi splitting of
exciton-polariton modes formed by bright excitons and a confined
optical mode of the cavity~\cite{Kavokin2007}.

The origin of Eqs.~(\ref{eq:Hb}) and (\ref{eq:Hd}) can be easily seen from
the exciton Hamiltonian written in the basis of linear x and y
polarizations. For example, for the bright excitons:
\begin{align}
\hat{\mathcal{H}}_{XY}&=\left[%
\begin{array}{cc}
E_b-\delta_b & 0 \\
0 & E_b+\delta_b%
\end{array}%
\right] \\
\hat{\mathcal{H}}_b&=\hat{C}^{-1}\hat{\mathcal{H}}_{XY}\hat{C},
\end{align}
where:
\begin{equation}
\hat{C}=\frac{1}{\sqrt{2}}\left[%
\begin{array}{cc}
1 & 1 \\
i & -i%
\end{array}%
\right],\hspace{10mm}\hat{C}^{-1}=\frac{1}{\sqrt{2}}\left[%
\begin{array}{cc}
1 & -i \\
1 & i%
\end{array}%
\right]
\end{equation}
are the transformation matrices from the linear to circular polarization
basis and vice versa~\cite{Born2012}. The same reasoning can be applied to
the dark excitons as well.

The sum of Hamiltonians $\hat{\mathcal{H}}_{b}$ and $\hat{\mathcal{H}}_{d}$,
written in the $4\times 4$ exciton spin basis is:
\begin{equation}
\hat{\mathcal{H}}_{ex}^{ex}=\left[
\begin{array}{cccc}
E_{b} & -\delta _{b} & 0 & 0 \\
-\delta _{b} & E_{b} & 0 & 0 \\
0 & 0 & E_{d} & -\delta _{d} \\
0 & 0 & -\delta _{d} & E_{d}%
\end{array}%
\right]  \label{eq:H04x4}
\end{equation}

Now, the full exciton Hamiltonian can be written as:

\begin{widetext}
\begin{align}
\hat{\mathcal{H}}_{ex}^{tot}&=\hat{\mathcal{H}}_{ex}^{e}+\hat{\mathcal{H}}%
_{ex}^{h}\hat{+\mathcal{H}}_{ex}^{ex}\notag\\&=\left[
\begin{array}{cccc}
E_{b}+\frac{1}{2}\left( g_{e}-g_{h}\right) \mu _{B}B & -\delta _{b} & \beta
_{e}k_{e}e^{-i\phi } & \beta _{h}k_{h}e^{-i\phi } \\
-\delta _{b} & E_{b}-\frac{1}{2}\left( g_{e}-g_{h}\right) \mu _{B}B & \beta
_{h}k_{h}e^{i\phi } & \beta _{e}k_{e}e^{i\phi } \\
\beta _{e}k_{e}e^{i\phi } & \beta _{h}k_{h}e^{-i\phi } & E_{d}-\frac{1}{2}%
\left( g_{e}+g_{h}\right) \mu _{B}B & -\delta _{d} \\
\beta _{h}k_{h}e^{i\phi } & \beta _{e}k_{e}e^{-i\phi } & -\delta _{d} &
E_{d}+\frac{1}{2}\left( g_{e}+g_{h}\right) \mu _{B}B%
\end{array}%
\right].\label{eq:H4x4}
\end{align}
\end{widetext}

For the translational motion of an exciton as a whole particle the exciton
momentum is given by $\mathbf{P}_{ex}=\left( m_{e}+m_{h}\right) \mathbf{v}%
_{ex}$ , where $m_{e}$ and $m_{h}$ are in-plane effective masses of an
electron and of a heavy hole, respectively; $\mathbf{v}_{ex}$ is the exciton
velocity. Having in mind that the exciton translational momentum is a sum of
electron and hole translational momenta given by $\mathbf{P}_{\mathrm{e,h}%
}=m_{e,h}\mathbf{v}_{e,h}$, where $\mathbf{v}_{e}$ and $\mathbf{v}_{h}$ are
the electron and hole velocity, respectively, one can easily see that $%
\mathbf{v}_{h}=\mathbf{v}_{e}=\mathbf{v}_{ex}$. Having in mind that $\mathbf{%
P}_{ex}=\hbar \mathbf{k}_{ex}$ and $\mathbf{P}_{e,h}=\hbar \mathbf{k}_{e,h}$%
, we have $\mathbf{k}_{ex}=\mathbf{k}_{h}+\mathbf{k}_{e}$, with $\mathbf{k}%
_{e}=\frac{m_{e}}{m_{e}+m_{h}}\mathbf{k}_{ex}$ and $\mathbf{k}_{h}=\frac{%
m_{h}}{m_{e}+m_{h}}\mathbf{k}_{ex}$. Thus, $\hat{\mathcal{H}}_{ex}^{tot}$
depends on the exciton center of mass wave-vector $\mathbf{k}_{ex}$ and on
the angle $\phi $ between this angle and one of the structure axes (e.g.
(100)-axis).

It should be noted that in this consideration the wave-vectors $\mathbf{k}%
_{ex},\mathbf{k}_{h},\mathbf{k}_{e}$ are related to the translational motion
of the exciton as a whole particle, with hole and electron as its
constituents. The wave-vector of relative motion of the electron and hole
``inside'' the exciton is zero on average but may be important for each
given moment of time. Recently, the effect of relative electron-hole motion
on the spin-orbit effects of excitons has been analyzed by Vishnevsky
\textit{et al} \cite{Vishnevski} . Their analysis confirms the presence of
linear in $\mathbf{k}_{ex}$ spin orbit terms in the exciton Hamiltonian
introduced above.

\subsection{Spin Density Matrix}

\label{sec:DensityMatrix} Having constructed the Hamiltonian for excitons
propagating with a wavevector $\mathbf{k}_\mathrm{ex}$, we now consider the
description of their spin state. We shall use the spin density matrix, $\hat{%
\rho}=\left|\Psi\right>\left<\Psi\right|$, where $\Psi=\left(\Psi_{+1},%
\Psi_{-1},\Psi_{+2},\Psi_{-2}\right)$ are the components of the exciton
wavefunction projected onto the four spin states, $\left(\left|\Psi_{+1}%
\right>,\left|\Psi_{-1}\right>,\left|\Psi_{+2}\right>,\left|\Psi_{-2}\right>%
\right)$.

\subsubsection{Relation to Stokes' vectors and polarization degrees of light}

The exciton spin DM is given by:
\begin{align}
\hat{\rho}& =\left\vert \Psi \right\rangle \left\langle \Psi \right\vert
\notag \\
& =\left[
\begin{array}{cccc}
\Psi _{+1}^{\ast }\Psi _{+1} & \Psi _{-1}^{\ast }\Psi _{+1} & \Psi
_{+2}^{\ast }\Psi _{+1} & \Psi _{-2}^{\ast }\Psi _{+1} \\
\Psi _{+1}^{\ast }\Psi _{-1} & \Psi _{-1}^{\ast }\Psi _{-1} & \Psi
_{+2}^{\ast }\Psi _{-1} & \Psi _{-2}^{\ast }\Psi _{-1} \\
\Psi _{+1}^{\ast }\Psi _{+2} & \Psi _{-1}^{\ast }\Psi _{+2} & \Psi
_{+2}^{\ast }\Psi _{+2} & \Psi _{-2}^{\ast }\Psi _{+2} \\
\Psi _{+1}^{\ast }\Psi _{-2} & \Psi _{-1}^{\ast }\Psi _{-2} & \Psi
_{+2}^{\ast }\Psi _{-2} & \Psi _{-2}^{\ast }\Psi _{-2},%
\end{array}%
\right]
\end{align}%
The elements of the upper left quarter of this DM are linked to the
intensity of light emitted by bright exciton states, $I=\Psi _{+1}^{\ast
}\Psi _{+1}+\Psi _{-1}^{\ast }\Psi _{-1}$, and to the components of the
Stokes' vector, $S_{x}$, $S_{y}$ and $S_{z}$ of the emitted light:
\begin{align}
\rho _{11}& =\frac{I}{2}+S_{z}, \\
\rho _{12}& =S_{x}-iS_{y}, \\
\rho _{21}& =S_{x}+iS_{y}, \\
\rho _{22}& =\frac{I}{2}-S_{z}.
\end{align}%
These expressions can be summarized more succinctly using the Pauli matrices
as:
\begin{equation}
\left[
\begin{array}{cc}
\rho _{11} & \rho _{12} \\
\rho _{21} & \rho _{22}%
\end{array}%
\right] =\frac{I}{2}\hat{I}+\mathbf{S}.\mathbf{\hat{\sigma}},
\end{equation}%
where $\mathbf{S}=\left( S_{x},S_{y},S_{z}\right) $ is the Stokes' vector
and we recall that $\hat{I}$ is the identity matrix.
{
{Note that the trace of the spin density matrix is a
number of particles in the system, which is not conserved because of
the finite lifetime, in contrast with the full quantum optical
density matrix which has the trace equal to unity.}}

Often when studying the polarization structure of fields with non-uniform
intensity, it is useful to compare the polarization degrees of emitted
light, which can be given by normalizing the Stokes' vectors to the light
intensity. The circular polarization degree is:
\begin{equation}
\rho_c=\frac{2S_z}{I}=\frac{\rho_{11}-\rho_{22}}{\rho_{11}+\rho_{22}}.
\end{equation}
The horizontal-vertical linear polarization degree is:
\begin{equation}
\rho_l=\frac{2S_x}{I}=\frac{\rho_{12}+\rho_{21}}{\rho_{11}+\rho_{22}}.
\end{equation}
The linear polarization degree measured in the diagonal axes (also referred
to as a diagonal polarization degree) is given by:
\begin{equation}
\rho_d=\frac{2S_y}{I}=i\frac{\rho_{12}-\rho_{21}}{\rho_{11}+\rho_{22}}.
\end{equation}

\subsubsection{Liouville equation}

The dynamics of the DM is given by the quantum Liouville equation:
\begin{equation}
i\hbar\frac{d\hat{\rho}}{dt}=\left[ \hat{\mathcal{H}}_{ex}^{tot},\hat{\rho}%
\right] ,  \label{eq:Liouville1}
\end{equation}%
where the Hamiltonian is composed from the electron, hole and exciton
contributions given by Eqs.~(\ref{eq:He4x4}), (\ref{eq:Hh4x4}) and (\ref%
{eq:H04x4}) (considering the Faraday magnetic field configuration).

So far, we have neglected all relaxation or scattering processes in the
system. The commonly used way to account for these processes is through the
introduction of a phenomenological Lindblad superoperator to the Liouville
equation:
\begin{equation}
i\hbar \frac{d\hat{\rho}}{dt}=\left[ \hat{\mathcal{H}}_{ex}^{tot},\hat{\rho}%
\right] -\hat{L}\left( \hat{\rho}\right) ,  \label{eq:Liouville}
\end{equation}%
where the Lindblad superoperator is introduced as:
\begin{equation}
\hat{L}\left( \hat{\rho}\right) =i\hbar \left[
\begin{array}{cccc}
\rho _{11}/\tau _{b} & \rho _{12}/\tau _{b} & \rho _{13}/\tau _{c} & \rho
_{14}/\tau _{c} \\
\rho _{21}/\tau _{b} & \rho _{22}/\tau _{b} & \rho _{23}/\tau _{c} & \rho
_{24}/\tau _{c} \\
\rho _{31}/\tau _{c} & \rho _{32}/\tau _{c} & \rho _{33}/\tau _{d} & \rho
_{34}/\tau _{d} \\
\rho _{41}/\tau _{c} & \rho _{42}/\tau _{c} & \rho _{43}/\tau _{d} & \rho
_{44}/\tau _{d}%
\end{array}%
\right] .
\end{equation}%
$\tau _{b}$ is the bright exciton decoherence time, $\tau _{d}$ is the dark
exciton decoherence time and $\tau _{c}$ is the characteristic decoherence
time of processes between dark and bright excitons. Note that dissipation
may be crucial in the description of exciton spin currents in realistic
systems. In particular, within this formalism, in the presence of
dissipation, the current conservation and flux conservation conditions
become valid if completed by exciton generation and decay in the continuity
equation. In the rest of this manuscript we shall neglect dissipation to
simplify the model system and to clarify the physical mechanisms which
govern the characteristics of exciton spin currents. Namely, we shall assume
$\hat{L}\left( \hat{\rho}\right) =0$. We stress that, in the experiment, the
magnitude of predicted spin currents may be reduced by dissipation.


The formalism described so far, in sections~\ref{sec:Hamiltonian} and this
section (\ref{sec:DensityMatrix}), has been successfully applied in the
description of spin transport in gases of cold excitons in coupled
GaAs/AlGaAs quantum wells~\cite{High2013}. In this work, cold excitons are
generated within localized spots and then fly away ballistically in radial
directions. The elements of the DM, $\rho _{ij}$, are dependent on the
distance from the excitation spot $\mathbf{r}=\mathbf{v}_{ex}t$ at time $t$
and the polar angle, $\phi $. The propagation speed $\mathbf{v}_{ex}=\hbar
\mathbf{k}_{ex}/\left( m_{e}+m_{h}\right) $. In this work, when solving the
Liouville equation introduced above, we shall refer to the experimental
configuration of Ref.~\onlinecite{High2013}. In particular, this implies a
specific choice of the initial conditions for Eq. ~(\ref{eq:Liouville1}): we
shall assume that at zero time excitons are not moving. We shall assume that
they populate the eigenstates of the exciton Hamiltonian $\hat{\mathcal{H}}%
_{ex}^{tot}$ taken with $\mathbf{k}_{ex}=0$ following a thermal
distribution with a temperature $T$. We shall assume that once
created in the equilibrium state, the excitons start moving apart in
the radial direction. Thus, implicitly, we account for a non-linear
effect: dipole-dipole repulsion of
excitons which makes them acquire a certain in-plane velocity $\mathbf{v}%
_{ex}$. This non-linearity is crucial to move the system out of equilibirum.
The rest of exciton propagation and spin dynamics is modelled using the
linear equation (\ref{eq:Liouville1}), where the Hamiltonian $\hat{\mathcal{H%
}}_{ex}^{tot}$ contains now the off-diagonal terms proportional to $\mathbf{%
k}_{ex}$. 

\subsection{Electron and hole spin currents}

\label{sec:ehcurrents} We shall normalize exciton, electron and heavy-hole
functions to unity, namely:
\begin{align}
\Psi _{+1}^{\ast }\Psi _{+1}+\Psi _{-1}^{\ast }\Psi _{-1}+\Psi _{+2}^{\ast
}\Psi _{+2}+\Psi _{-2}^{\ast }\Psi _{-2}& =1 \\
\Psi _{e,+\frac{1}{2}}^{\ast }\Psi _{e,+\frac{1}{2}}+\Psi _{e,-\frac{1}{2}%
}^{\ast }\Psi _{e,-\frac{1}{2}}& =1 \\
\Psi _{h,+\frac{3}{2}}^{\ast }\Psi _{h,+\frac{3}{2}}+\Psi _{h,-\frac{3}{2}%
}^{\ast }\Psi _{h,-\frac{3}{2}}& =1
\end{align}

Now, the exciton spin DM (21) can be represented in terms of electron and
hole wavefunctions as:
\begin{widetext}
\begin{equation}
\hat{\rho}=\left[\begin{array}{cccc}
\Psi^*_{e,-\frac{1}{2}}\Psi_{e,-\frac{1}{2}}\Psi^*_{h,+\frac{3}{2}}
\Psi_{h,+\frac{3}{2}}&\Psi^*_{e,+\frac{1}{2}}\Psi_{e,-\frac{1}{2}}
\Psi^*_{h,-\frac{3}{2}}\Psi_{h,+\frac{3}{2}}&\Psi^*_{e,+\frac{1}{2}}
\Psi_{e,-\frac{1}{2}}\Psi^*_{h,+\frac{3}{2}}\Psi_{h,+\frac{3}{2}}&
\Psi^*_{e,-\frac{1}{2}}\Psi_{e,-\frac{1}{2}}\Psi^*_{h,-\frac{3}{2}}
\Psi_{h,+\frac{3}{2}}\\
\Psi^*_{e,-\frac{1}{2}}\Psi_{e,+\frac{1}{2}}\Psi^*_{h,+\frac{3}{2}}
\Psi_{h,-\frac{3}{2}}&\Psi^*_{e,+\frac{1}{2}}\Psi_{e,+\frac{1}{2}}
\Psi^*_{h,-\frac{3}{2}}\Psi_{h,-\frac{3}{2}}&\Psi^*_{e,+\frac{1}{2}}
\Psi_{e,+\frac{1}{2}}\Psi^*_{h,+\frac{3}{2}}\Psi_{h,-\frac{3}{2}}&
\Psi^*_{e,-\frac{1}{2}}\Psi_{e,+\frac{1}{2}}\Psi^*_{h,-\frac{3}{2}}
\Psi_{h,-\frac{3}{2}}\\
\Psi^*_{e,-\frac{1}{2}}\Psi_{e,+\frac{1}{2}}\Psi^*_{h,+\frac{3}{2}}
\Psi_{h,+\frac{3}{2}}&\Psi^*_{e,+\frac{1}{2}}\Psi_{e,+\frac{1}{2}}
\Psi^*_{h,-\frac{3}{2}}\Psi_{h,+\frac{3}{2}}&\Psi^*_{e,+\frac{1}{2}}
\Psi_{e,+\frac{1}{2}}\Psi^*_{h,+\frac{3}{2}}\Psi_{h,+\frac{3}{2}}&
\Psi^*_{e,-\frac{1}{2}}\Psi_{e,+\frac{1}{2}}\Psi^*_{h,-\frac{3}{2}}
\Psi_{h,+\frac{3}{2}}\\
\Psi^*_{e,-\frac{1}{2}}\Psi_{e,-\frac{1}{2}}\Psi^*_{h,+\frac{3}{2}}
\Psi_{h,-\frac{3}{2}}&\Psi^*_{e,+\frac{1}{2}}\Psi_{e,-\frac{1}{2}}
\Psi^*_{h,-\frac{3}{2}}\Psi_{h,-\frac{3}{2}}&\Psi^*_{e,+\frac{1}{2}}
\Psi_{e,-\frac{1}{2}}\Psi^*_{h,+\frac{3}{2}}\Psi_{h,-\frac{3}{2}}&
\Psi^*_{e,-\frac{1}{2}}\Psi_{e,-\frac{1}{2}}\Psi^*_{h,-\frac{3}{2}}
\Psi_{h,-\frac{3}{2}},
\end{array}\right]
\end{equation}
\end{widetext}This representation allows us to obtain useful links between
the elements of exciton, electron and hole density matrices, in particular:
\begin{align}
\hat{\rho}_{e}=\left\vert \Psi _{e}\right\rangle \left\langle \Psi
_{e}\right\vert & =\left[
\begin{array}{cc}
\Psi _{e,+\frac{1}{2}}^{\ast }\Psi _{e,+\frac{1}{2}} & \Psi _{e,-\frac{1}{2}%
}^{\ast }\Psi _{e,+\frac{1}{2}} \\
\Psi _{e,+\frac{1}{2}}^{\ast }\Psi _{e,-\frac{1}{2}} & \Psi _{e,-\frac{1}{2}%
}^{\ast }\Psi _{e,-\frac{1}{2}}%
\end{array}%
\right]  \notag \\
& =\left[
\begin{array}{cc}
\rho _{22}+\rho _{33} & \rho _{24}+\rho _{31} \\
\rho _{13}+\rho _{42} & \rho _{11}+\rho _{44}%
\end{array}%
\right] \\
\hat{\rho}_{h}=\left\vert \Psi _{h}\right\rangle \left\langle \Psi
_{h}\right\vert & =\left[
\begin{array}{cc}
\Psi _{h,+\frac{3}{2}}^{\ast }\Psi _{h,+\frac{3}{2}} & \Psi _{h,-\frac{3}{2}%
}^{\ast }\Psi _{h,+\frac{3}{2}} \\
\Psi _{h,+\frac{3}{2}}^{\ast }\Psi _{h,-\frac{3}{2}} & \Psi _{h,-\frac{3}{2}%
}^{\ast }\Psi _{h,-\frac{3}{2}}%
\end{array}%
\right]  \notag \\
& =\left[
\begin{array}{cc}
\rho _{11}+\rho _{33} & \rho _{14}+\rho _{32} \\
\rho _{23}+\rho _{41} & \rho _{22}+\rho _{44}%
\end{array}%
\right]
\end{align}%
We know that the components of electron and hole density matrices are linked
with the projections of electron and hole spins as:
\begin{align}
\hat{\rho}_{e}& =\left[
\begin{array}{cc}
\frac{1}{2}+S_{e,z} & S_{e,x}-iS_{e,y} \\
S_{e,x}+iS_{e,y} & \frac{1}{2}-S_{e,z}%
\end{array}%
\right] \\
\hat{\rho}_{h}& =\left[
\begin{array}{cc}
\frac{1}{2}+S_{h,z} & S_{h,x}-iS_{h,y} \\
S_{h,x}+iS_{h,y} & \frac{1}{2}-S_{h,z}%
\end{array}%
\right] ,
\end{align}%
where, for the heavy hole we have assigned spin +1/2 to the state +3/2 and
spin -1/2 to the state -3/2 accounting for the orbital momentum of these
states of +1 and -1, respectively.

The z-component of the spin polarization carried by electrons can now be
expressed as:
\begin{equation}
S_{e,z}=\left(\rho_{22}+\rho_{33}-\rho_{11}-\rho_{44}\right)/2.
\end{equation}
Similarly, the z-component of the spin polarization carried by holes can now
be expressed as:
\begin{equation}
S_{h,z}=\left(\rho_{11}+\rho_{33}-\rho_{22}-\rho_{44}\right)/2.
\end{equation}

The in-plane component of electron and hole spins can be extracted from the
off-diagonal elements of the DM. Namely, the x-component of electron spin is
given by:
\begin{equation}
S_{e,x}=\left(\rho_{13}+\rho_{31}+\rho_{24}+\rho_{42}\right)/2,
\end{equation}
while the x-component of the hole spin is given by:
\begin{equation}
S_{h,x}=\left(\rho_{14}+\rho_{23}+\rho_{32}+\rho_{41}\right)/2.
\end{equation}
The y-component of electron spin is given by:
\begin{equation}
S_{e,y}=i\left(-\rho_{13}+\rho_{31}+\rho_{24}-\rho_{42}\right)/2,
\end{equation}
while the y-component of the hole spin is given by:
\begin{equation}
S_{h,y}=i\left(\rho_{14}-\rho_{23}+\rho_{32}-\rho_{41}\right)/2,
\end{equation}

\section{Numerical Results in the Density Matrix formalism}

\begin{figure}[h]
\centering
\includegraphics[width=8.116cm]{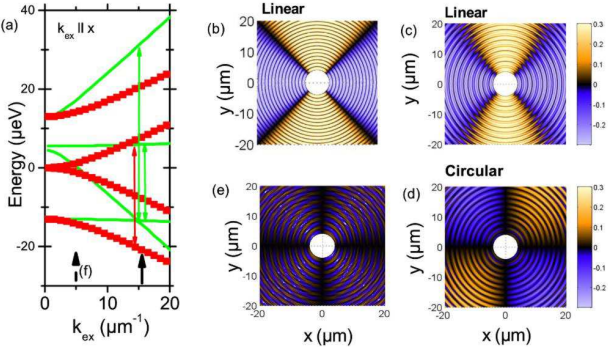}
\caption{(color online) (a) Dispersion of the excitonic states calculated
using the set of parameters from Table~\protect\ref{tab:param} (green lines)
and reduced set of parameters with $E_b-E_d=0$, $\protect\delta_b=0$, $%
\protect\beta_h=0$ (red symbols). Black arrows indicate the values of the
wavevectors used in (b)-(d) $k_{ex}=15.3 \protect\mu$m$^{-1}$ and (e) $%
k_{ex}=5 \protect\mu$m$^{-1}$. Double-ended arrows indicate the energies of
oscillations between the eigenstates which appear in spatial polarization
patterns. (b) Linear polarization degree along x-axis calculated with the
simplified set of parameters. (c) Same as (b) but for the full set of
parameters from the Table~\protect\ref{tab:param}. (d) Circular polarization
degree with parameters from Table~\protect\ref{tab:param}. (e) Same but at $%
k_{ex}=5 \protect\mu$m$^{-1}$. The source area was taken circular with a
radius of $4\protect\mu$m.}
\label{FigDisp}
\end{figure}
Figure~\ref{FigDisp} shows the numerical results obtained within the DM
formalism for a model system with the same parameters as those of coupled
double quantum wells studied in Ref.~\onlinecite{High2013}. The parameters
are summarized in Table~\ref{tab:param}.
\begin{table}[h]
\caption{Parameters for numerical calculations.}
\label{tab:param}\centering
\begin{tabular}{rcrl}
\hline
Electron mass & $m_e$ & $0.07$ & $m_0$ \\
Heavy hole mass & $m_h$ & $0.16$ & $m_0$ \\
Electron Dresselhaus coupling & $\beta_e$ & $2.7$ & $\mu$eV$\mu$m \\
Heavy-hole Dresselhaus coupling & $\beta_h$ & $0.92$ & $\mu$eV$\mu$m \\
Bright exciton XY splitting & $\delta_b$ & $0.5$ & $\mu$eV \\
Dark exciton XY splitting & $\delta_d$ & $-13$ & $\mu$eV \\
Bright-dark exciton splitting & $E_b-E_d$ & $5$ & $\mu$eV \\ \hline
\end{tabular}%
\end{table}

The dispersion of bright and dark exciton modes obtained by diagonalisation
of the Hamiltonian $\hat{\mathcal{H}}_{ex}^{tot}$ (20) is shown in Fig. \ref%
{FigDisp}(a) by green solid lines. The momentum has been chosen along the $x$
direction, but the anisotropy of the band structure remains small. The
initial splittings of dark and bright states makes these dispersion curves
qualitatively different from those presented by Vishnevsky \textit{et al }%
\cite{Vishnevski}. Note also, that Ref. (\cite{Vishnevski}) accounts for the
exciton kinetic energy which we neglect in $\hat{\mathcal{H}}_{ex}^{tot}$ ,
leaving only the spin-dependent contributions to the energy.
The numerically
calculated linear and circular polarization degrees are shown in Figs. \ref%
{FigDisp}(c) and (d), respectively.
{
{To stay close to the experimental conditions of
Ref. \cite{High2013}, we have chosen  as initial condition the
circular source area with a radius of $4$ $\mu$m, where cold
excitons are generated within localized spots.}}
In Fig.\ref{FigDisp}(c), four lobes are
unambiguously observable in the pattern of the linear and diagonal (not
shown) polarizations. This pattern is a consequence of the Dresselhaus
spin-orbit coupling for electrons, and it is characteristic of the chosen
initial state at the source: four split eigen-states with zero in-plane
wave-vector are occupied. Relative occupation of these states corresponds to
the Boltzmann distribution at temperature $T=0.1$ K. This choice of initial
conditions leads to a variety of polarization patterns observed in
experiment \cite{High2013}.
{
{In the absence of the damping this pattern is
periodic and infinite in the radial direction.}}

In order to reveal the mechanism of formation of the polarization patterns
it is instructive to consider a simplified version of the Hamiltonian~\ref%
{eq:H4x4}. In the absence of magnetic field, zero Dresselhaus effect for
holes, zero splitting of bright excitons and zero splitting $E_{b}-E_{d}$
between bright and dark excitons this Hamiltonian can be rewritten as:
\begin{equation}
\hat{\mathcal{H}}=-\delta _{d}\left[
\begin{array}{cccc}
0 & 0 & \xi e^{-i\phi } & 0 \\
0 & 0 & 0 & \xi e^{i\phi } \\
\xi e^{i\phi } & 0 & 0 & 1 \\
0 & \xi e^{-i\phi } & 1 & 0%
\end{array}%
\right]  \label{eq:HBL}
\end{equation}%
where $\xi =-\beta _{e}k_{e}/\delta _{d}$.
{
{In other words, here we only take into account
linear splitting of dark exciton states which  inevitably results
from structural anisotropy even in the best quality samples, and the
Dresselhaus field acting on the spin of electron, bound to the hole.
We will see that at sufficiently low temperature these two
ingredients provide the in-plane asymmetry that ultimately results
in the formation of linear polarisation patterns.
Indeed,}} the eigenvalues of this Hamiltonian can be obtained
analytically : $E=\pm \frac{1}{2}\delta _{d}\pm \frac{1}{2}\delta
_{d}\sqrt{1+4\xi ^{2}}$. This corresponds to two dispersion branches
at low energy $\propto \pm \xi ^{2}$ and two branches at high energy
$\propto \pm (1+\xi ^{2})$ for $\xi \ll 1$. These branches are shown
by red squares in Fig. \ref{FigDisp}(a).The eigenvectors, starting
from the lowest energy and taking $\delta _{d}<0$, can be
approximated for small $\xi $ by:
\begin{equation}
\left[
\begin{array}{c}
\xi e^{-i\phi } \\
-\xi e^{i\phi } \\
-1 \\
1%
\end{array}%
\right] ,\left[
\begin{array}{c}
e^{-i\phi } \\
e^{i\phi } \\
-\xi \\
-\xi%
\end{array}%
\right] ,\left[
\begin{array}{c}
e^{-i\phi } \\
-e^{i\phi } \\
\xi \\
-\xi%
\end{array}%
\right] ,\left[
\begin{array}{c}
\xi e^{-i\phi } \\
\xi e^{i\phi } \\
1 \\
1%
\end{array}%
\right] .  \label{eq:eigenBL}
\end{equation}%
As mentioned already, at low temperatures $k_{B}T\ll |\delta _{d}|$,
the lowest energy state with zero momentum is given by $[0,0,-1,1]$
and is a linearly polarized dark exciton. By linearly polarised dark
exciton we mean the dark state which has a dipole moment oriented in
a certain way, and which has a zero spin projection to the grows
axis of the structure. After an acceleration due to dipole
repulsion, this initial state is no longer an eigenstate and
oscillates as a function of time between the two eigenstates of the
Hamiltonian with which the initial state is not orthogonal, namely
the first and the third eigenstates listed in Eqs.~\ref{eq:eigenBL}.
Among these two states, only the third gives a significant
contribution to the observed polarization as it is essentially
\textquotedblleft bright\textquotedblright\ (has large projections
to $+1$ and $-1$ exciton states). The linear and diagonal
polarizations originating from this state are readily given by
$-\cos (2\phi )$ and $-\sin (2\phi )$ respectively. This reproduces
the essential features of the numerical results for the linear
polarization pattern, as one can see comparing the images calculated
with the reduced Hamiltonian (Fig. \ref{FigDisp}(b)) and the full
Hamiltonian (Fig. \ref{FigDisp}(c)). In the particular case
considered here (initial state formed essentially by linearly
polarized dark excitons) the Dresselhaus spin-orbit term for
electrons leads to formation of the linear polarization vortex: the
polarization plane is always perpendicular to the wave vector
direction. The linear polarization vortex has been observed
experimentally by High et al \cite{High2012, High2013}.
{
{Rapid oscillations in radial direction are due
periodical change in the occupation of mainly dark and mainly bright
states, indicated by double-ended arrows in Fig.
(\ref{FigDisp}(b)).}}

We should emphasize that with a proper reordering of the basis vectors, the
simplified Hamiltonian~\ref{eq:HBL} is analytically equivalent to the
Hamiltonian of bilayer graphene.
In its simplest expression, the Hamiltonian of bilayer graphene can be
written as \cite{McCann2012}:
\begin{equation}
\left[
\begin{array}{cccc}
0 & v_F k e^{-i\phi} & 0 & 0 \\
v_F k e^{i\phi } & 0 & t_\perp & 0 \\
0 & t_\perp & 0 & v_F k e^{-i\phi} \\
0 & 0 & v_F k e^{i \phi} & 0 \\
&  &  &
\end{array}%
\right] .  \label{eq:Hbilayer}
\end{equation}%
where $v_F$ is the velocity, $k$ the amplitude of the momentum, $\phi$ the
angle between the momentum and the $x$ axis, and $t_\perp$ is the main
coupling term between the two graphene layers, which is given by hopping
between two carbon atoms that are superimposed. The four coefficients of the
associated wavevectors correspond to the probability amplitudes on the two
independent sublattices of the two graphene layers. By permutation of the
basis vectors:$1 \rightarrow 1$, $2 \rightarrow 3$, $3 \rightarrow 4$, $4
\rightarrow 2$, the Hamiltonian is rewritten as:
\begin{equation}
\left[
\begin{array}{cccc}
0 & 0 & v_F k e^{-i\phi} & 0 \\
0 & 0 & 0 & v_F k e^{i\phi} \\
v_F k e^{i\phi} & 0 & 0 & t_\perp \\
0 & v_F k e^{-i \phi} & t_\perp & 0 \\
&  &  &
\end{array}%
\right] .
\end{equation}%
which is equivalent to Eq.~\ref{eq:HBL} with $t_\perp \equiv -\delta_d$ and $%
v_F k /t_\perp \equiv \xi$.

The $4\times4$ Hamiltonian of bilayer graphene is often restricted to the
subspace of the two low energy bands. From Eq.~\ref{eq:eigenBL}, the
corresponding eigenstates are:
\begin{equation}
[e^{i\phi}, \pm e^{-i\phi}].  \label{eq:H2x2}
\end{equation}
We can then define a pseudospin vector, which represents the relative phase
between the two components of the wavevectors. From Eq.~\ref{eq:H2x2}, the
pseudospin rotates two times when the particle wavevector undergoes one full
rotation.

Adapting the terminology used for graphene, the polarization pattern
observed in Fig. \ref{FigDisp}(b,c) would be nothing else than the
fingerprint of the ``pseudospin'' rotation in the exciton system. In the
context of the exciton system under consideration, the phase $\phi $
corresponds to the angle between the exciton polarization vector and the
chosen $x$-axis of the structure.


The build up of circular polarization requires introduction in the model of
the splitting between linearly polarized bright exciton states $\delta _{b}$
. This splitting acts as an effective magnetic field applied to the Stokes
vector of light emitted by bright excitons, $\mathbf{S}=(S_{x},S_{y},S_{z})$%
. In particular, if the $x$-polarized exciton state has a lower energy than
the $y$-polarized exciton state, it creates an effective magnetic field in
the $x$-direction which rotates the Stokes vector in the $yz$ plane. This
converts the diagonal polarization to the circular polarization and leads to
the appearance of right- and left-circularly polarized sectors in the
polarization map of exciton emission. Separation of spins due to
linear-to-circular polarization conversion is known in exciton-polariton
systems as the \textit{Optical spin Hall effect}. This effect was
theoretically predicted in Ref.~\onlinecite{Kavokin2005} and experimentally
observed in polaritonic \cite{Leyder2007} and excitonic systems \cite%
{High2013}. For microcavity polaritons it may be described in terms of beats
between TE and TM polarized polariton modes, while in the exciton system
studied here the effect is more complex due to the mixture of four nearly
degenerate (dark and bright) exciton states. A recent theoretical paper \cite%
{Vishnevski} predicts the skyrmion formation in this case. Here we
concentrate on the circular polarization patterns appearing due to the beats
between linearly polarized exciton states mixed by the exchange interaction.
Note that the circular polarization pattern is strongly sensitive to the
chosen exciton wave-vector, which governs the energies of the four involved
eigen-states. Figure~\ref{FigDisp}(d) is calculated assuming $k_{ex}=15.3\mu
m^{-1}$. This corresponds to the crossing point of the dispersion branches
associated to the first and second exciton eigen-states (see Fig \ref%
{FigDisp}(a)). For comparison, Fig.~\ref{FigDisp}(e) shows the circular
polarization pattern calculated with $k_{ex}=5\mu m^{-1}$. One can see that
the four lobes pattern of circular polarization is washed out, because the
lowest energy state of the system remains essentially dark. Rapid
oscillations that show up, have the period determined by the splitting
between the two lowest states.

Finally, let us underline that the dispersion curves shown in Fig. \ref%
{FigDisp}(a) do not take into account the kinetic energy of excitons. The
kinetic energy would shift all curves up by $K=\frac{\hbar ^{2}k_{ex}^{2}}{%
2m_{ex}}$. Note that this does not affect the splittings between exciton
eigenstates and would not affect the spin dynamics of excitons. In the spin
density formalism developed above we assign to all excitons the same kinetic
energy, $K$. In a realistic system, the kinetic energy may be spread, in
which case averaging of the obtained polarization patterns over $k_{ex}$ may
be needed. This averaging would smooth the fast oscillations seen on the
images Fig. \ref{FigDisp}~(c,d,e,f). In the next section devoted to the
non-linear spin dynamics we will fully take into account the kinetic energy
of propagating excitons.

\begin{figure}[h]
\centering
\includegraphics[width=10cm]{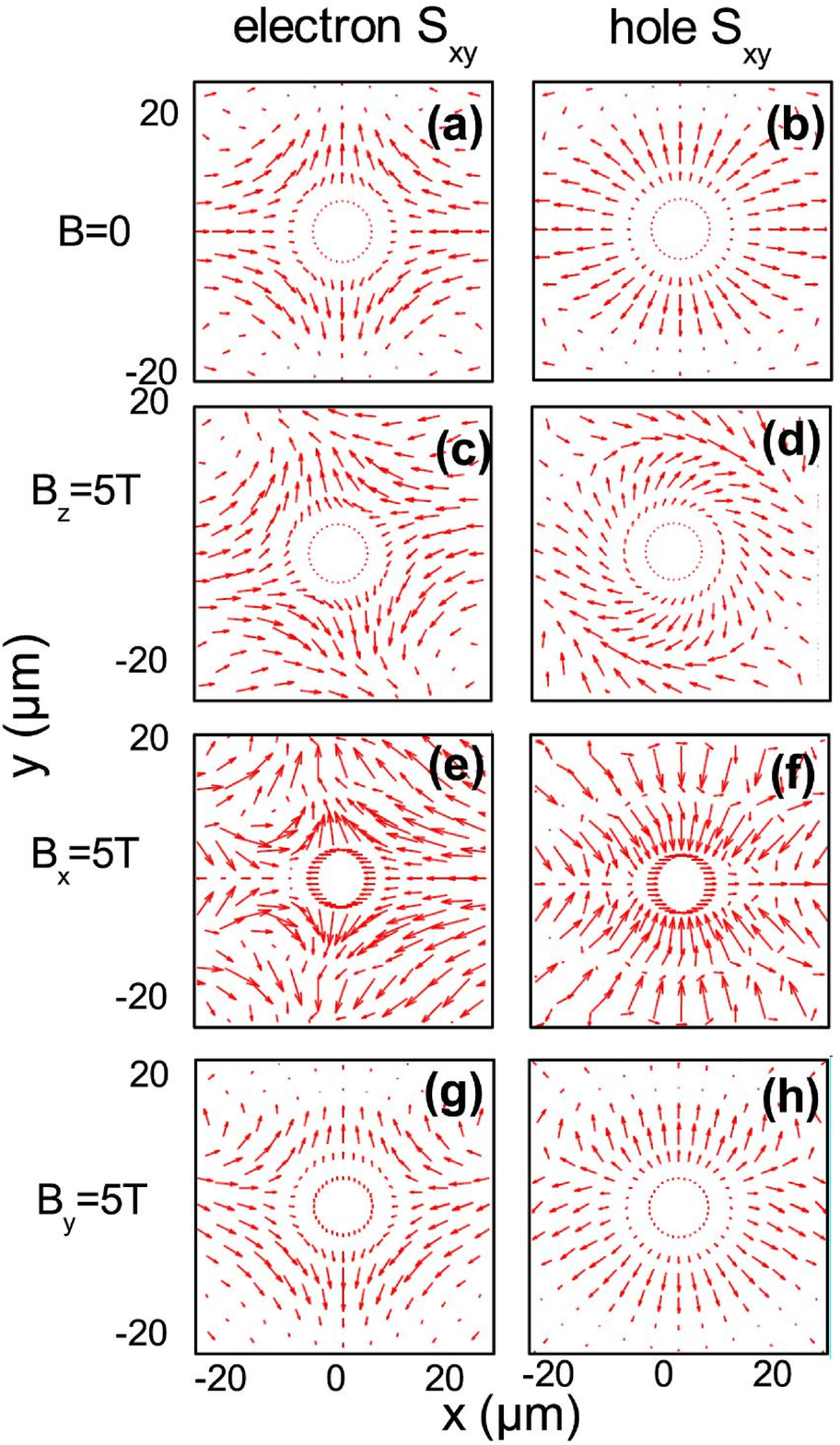}
\caption{(color online) Spatial distribution of electron and hole spin
components in the plane of the quantum well structure, calculated at zero
magnetic field and at $B=5$ T at three different orientations. The
parameters are given in Table~\protect\ref{tab:param}, at $\mathbf{k}%
_{ex}=15.3\protect\mu$m$^{-1}$ .The source area was taken circular with a
radius of $4\protect\mu$m.}
\label{FigSeh}
\end{figure}

The distribution of in-plane projections of electron and hole spins
for the same choice of parameters as above is shown in
Figs.~\ref{FigSeh}(a-h). The left panels show electron spins and the
right panels show the hole spins.
The direction of in-plane spin component is shown by arrows, while
the length of each arrow is proportional to the computed value of
the transverse spin component.
%
The upper panels, Figs.~\ref{FigSeh}(a,b), show the spin
distributions in the absence of a magnetic field. In this case the
electron and hole spins are oriented along the effective Dresselhaus
fields which are oriented differently for electrons and heavy holes,
as we have discussed in the previous section.
The decrease of the in-plane spin component upon propagation
corresponds to the build-up of the $z$-component of electron and
hole spins (Figure not shown), due to rotation of the exciton spin
around the effective magnetic field.
The magnetic field strongly changes the spin
distribution in real space. The spin textures become strongly
anisotropic in the case of in-plane (x- or y-oriented) magnetic
field. Note, that the
in-plane isotropy in the system is broken by the splitting between $x$- and $%
y$- polarized exciton states, which is why switching of the magnetic field
between $x$ and $y$ axes strongly affects the distribution of electron and
hole spins. It should be noted also that electron and hole in-plane spin
textures can hardly be observed directly in optical experiments. However,
they can be deduced from fitting the exciton polarization maps, e.g., using
the formalism described above.

\begin{figure}[h]
\centering
\includegraphics[width=8.116cm]{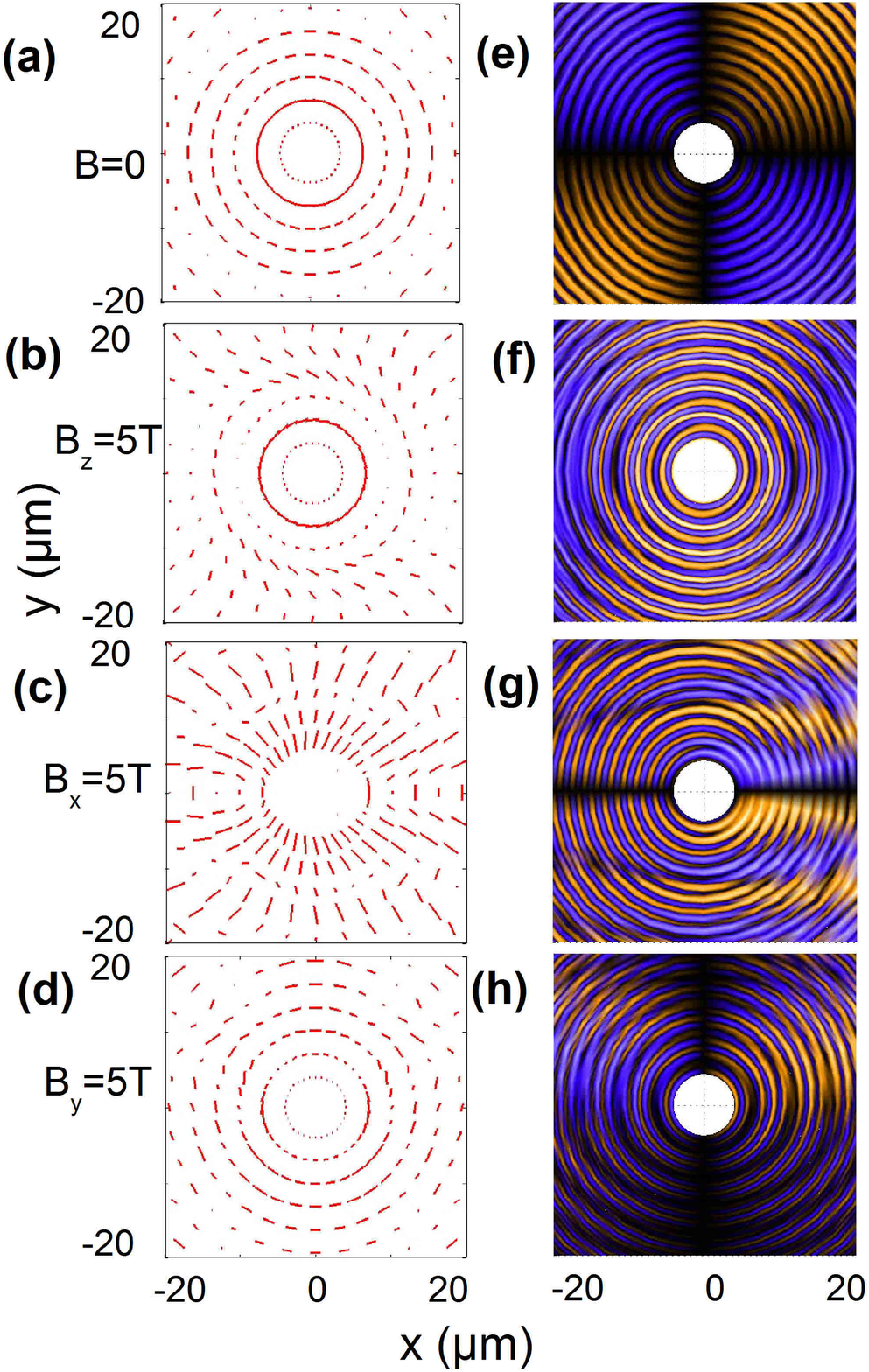}
\caption{(color online) Spatial distribution of exciton linear polarization
((a)-(d), length of the bars maps the intensity in arbitrary units) and
color maps of the exciton circular polarization degree ((e)-(f)), same color
code as in Fig. \protect\ref{FigDisp}). Magnetic field $B=0$ in (a), (e),
and $B=5$T along $z$ in (b),(f), $x$ in (c),(g), $y$ in (d),(h). Parameters
are given in Table~\protect\ref{tab:param}, $\mathbf{k}_{ex}=15.3\protect\mu$%
m$^{-1}$. The source area was taken circular with a radius of $4\protect\mu$%
m. }
\label{Fig_PolarB}
\end{figure}

Figure~\ref{Fig_PolarB} shows how the magnetic field affects spatial
patterns of linear (a-d) and circular (e-h) polarization. Switching the
magnetic field orientation between the $x$-, $y$- and $z$-axes one can
dramatically affect the polarization patterns. Having in mind that the
exciton polarization patterns can be directly observed in near-field
photoluminescence experiments, fitting of these patterns to the experimental
data would allow extracting the Dresselhaus constants and exciton exchange
splittings, which, in turn, allow to restore electron and hole spin textures
\cite{High2013}.

\begin{figure}[h]
\centering
\includegraphics[width=8.116cm]{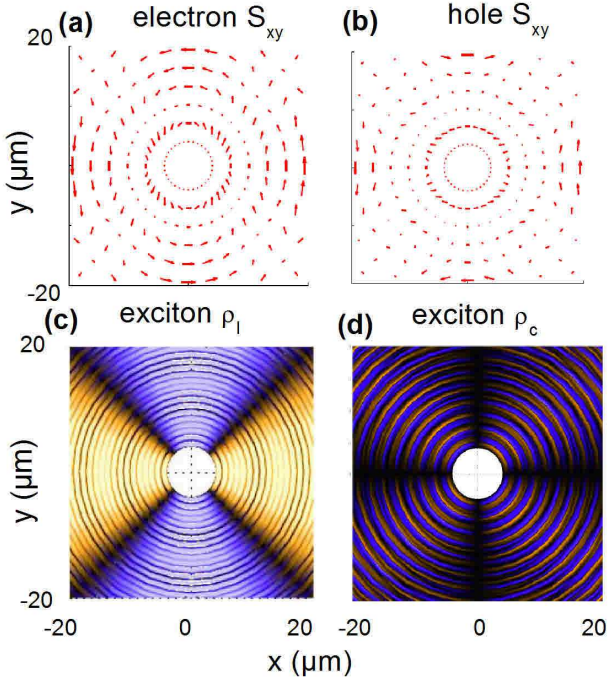}
\caption{(color online) Spatial distribution of electron (a) and hole (b)
in-plane spin component calculated at B=0 in the absence of the Dresselhaus
field $\protect\beta_e=0$; $\protect\beta_h=0$, but including Rashba field $%
\protect\alpha_e=2.7\protect\mu$eV$\protect\mu$m;. Other parameters are
given in Table~\protect\ref{tab:param}, $\mathbf{k}_{ex}=15.3\protect\mu$m$%
^{-1}$. The source area was taken circular with a radius of $4\protect\mu$m.
Corresponding patterns of linear (c) and circular (d) exciton polarization
degree are shown with the same color code as in Fig. \protect\ref{FigDisp}. }
\label{FigRashba}
\end{figure}

Figure \ref{FigRashba}(a-d) illustrates a peculiar regime where the
Dresselhaus fields for electrons and holes are taken to be zero and there is
no magnetic field applied, but electrons are subjected to the Rashba field
(the Rashba field for heavy holes is zero). Figure~\ref{FigRashba}(a) shows
the electron spin distribution in space, where the spins are clearly aligned
along the Rashba field force lines. Interestingly, the hole spins become
polarized as well, while no effective field acts on them (Figure \ref%
{FigRashba}(b)). This is an illustration of the exciton effect: bound in
excitons by Coulomb interaction and subject to the exchange induced exciton
effects, the holes acquire in-plane spin polarization. The non-zero spin
polarization of heavy holes is possible due to the exciton exchange effects.
Figure \ref{FigRashba}(c,d) shows the resulting linear and circular exciton
polarization patterns. One can see that the Rashba effect induces
polarization patterns strongly different from those induced by the
Dresselhaus effect, which is why from the shape of polarization patterns one
can conclude on the nature of spin-orbit coupling in the system.

\section{Non-linear spin dynamics of propagating excitons and
exciton-polaritons}

\label{sec:GP} In the previous section we operated with a spin density
matrix which is very convenient for the description of partially coherent
and partially polarized exciton gases. The quantum Liouville equation (\ref%
{eq:Liouville}) is a very efficient tool for the description of
effects linear in the exciton density. On the other hand, one cannot
straightforwardly incorporate non-linear interaction terms in this
equation. The treatment of non-linear effects in a partially
coherent system is a non-trivial task. Much simpler is the treatment
of non-linear effects in a perfectly coherent system, such as a
condensate at zero temperature. In this case, the ensemble of
excitons can be described by a single 4-component wave function
$\Psi=\left(\Psi_{+1},\Psi_{-1},\Psi_{+2},\Psi_{-2}\right)^T$. The
linear dynamics of this wave-function for ballistically propagating
excitons having a wavevector $\mathbf{k}_{ex}$ is described by the
Schr\"odinger equation:
\begin{equation}
i\hbar\frac{d}{dt}\left|\Psi\right>=\hat{\mathcal{H}}\left|\Psi\right>,
\label{eq:Shrodinger}
\end{equation}
where the Hamiltonian is the same as in Eq.~(\ref{eq:H4x4}). This equation
represents a set of four coupled linear differential equations for four
exciton spin components. Non-linear effects lead to the condensate evolution
in real and reciprocal space.

From now on we shall consider the exciton spin dynamics in real space (2D),
so that the wave function $\Psi$ will become coordinate-dependent and will
not be restricted to one single value of $\mathbf{k}_{ex}$. The non-linear
interaction terms for multi-component exciton gases are introduced and
discussed in detail in Ref.~\onlinecite{Rubo2011}. Here we expand Eq.~(\ref%
{eq:Shrodinger}) by introducing the kinetic energy (to describe the real
space dynamics) and the interaction terms. On the other hand, we neglect the
magnetic field, for simplicity. This results in a system of four non-linear
Schr\"odinger or GP equations~\cite{Rubo2011,Matuszewski2012,Kyriienko2012}:
\begin{align}
i\hbar\frac{d\Psi_{+1}}{dt}&=-\frac{\hbar^2\hat{\nabla}^2}{2m_{ex}}\Psi_{+1}+%
\frac{\beta_em_e}{m_{ex}}\left(\hat{k}_x-i\hat{k}_y\right)\Psi_{+2}  \notag
\\
&\hspace{5mm}+\frac{\beta_hm_h}{m_{ex}}\left(\hat{k}_x-i\hat{k}%
_y\right)\Psi_{-2}+\alpha_1\left|\Psi_{+1}\right|^2\Psi_{+1}  \notag \\
&\hspace{5mm}+\alpha_2\left|\Psi_{-1}\right|^2\Psi_{+1}+\alpha_3\left|%
\Psi_{+2}\right|^2\Psi_{+1}  \notag \\
&\hspace{5mm}+\alpha_4\left|\Psi_{-2}\right|^2\Psi_{+1}+W\Psi^*_{-1}%
\Psi_{+2}\Psi_{-2},  \label{eq:GP1}
\end{align}
\begin{align}
i\hbar\frac{d\Psi_{-1}}{dt}&=-\frac{\hbar^2\hat{\nabla}^2}{2m_{ex}}\Psi_{-1}+%
\frac{\beta_em_e}{m_{ex}}\left(\hat{k}_x+i\hat{k}_y\right)\Psi_{-2}  \notag
\\
&\hspace{5mm}+\frac{\beta_hm_h}{m_{ex}}\left(\hat{k}_x+i\hat{k}%
_y\right)\Psi_{+2}+\alpha_1\left|\Psi_{-1}\right|^2\Psi_{-1}  \notag \\
&\hspace{5mm}+\alpha_2\left|\Psi_{+1}\right|^2\Psi_{-1}+\alpha_3\left|%
\Psi_{-2}\right|^2\Psi_{-1}  \notag \\
&\hspace{5mm}+\alpha_4\left|\Psi_{+2}\right|^2\Psi_{-1}+W\Psi^*_{+1}%
\Psi_{+2}\Psi_{-2},  \label{eq:GP2}
\end{align}
\begin{align}
i\hbar\frac{d\Psi_{+2}}{dt}&=-\frac{\hbar^2\hat{\nabla}^2}{2m_{ex}}\Psi_{+2}+%
\frac{\beta_em_e}{m_{ex}}\left(\hat{k}_x+i\hat{k}_y\right)\Psi_{+1}  \notag
\\
&\hspace{5mm}+\frac{\beta_hm_h}{m_{ex}}\left(\hat{k}_x-i\hat{k}%
_y\right)\Psi_{-1}+\alpha_1\left|\Psi_{+2}\right|^2\Psi_{+2}  \notag \\
&\hspace{5mm}+\alpha_2\left|\Psi_{-2}\right|^2\Psi_{+2}+\alpha_3\left|%
\Psi_{+1}\right|^2\Psi_{+2}  \notag \\
&\hspace{5mm}+\alpha_4\left|\Psi_{-1}\right|^2\Psi_{+2}+W\Psi^*_{-2}%
\Psi_{+1}\Psi_{-1},  \label{eq:GP3}
\end{align}
\begin{align}
i\hbar\frac{d\Psi_{-2}}{dt}&=-\frac{\hbar^2\hat{\nabla}^2}{2m_{ex}}\Psi_{-2}+%
\frac{\beta_em_e}{m_{ex}}\left(\hat{k}_x-i\hat{k}_y\right)\Psi_{-1}  \notag
\\
&\hspace{5mm}+\frac{\beta_hm_h}{m_{ex}}\left(\hat{k}_x+i\hat{k}%
_y\right)\Psi_{+1}+\alpha_1\left|\Psi_{-2}\right|^2\Psi_{-2}  \notag \\
&\hspace{5mm}+\alpha_2\left|\Psi_{+2}\right|^2\Psi_{-2}+\alpha_3\left|%
\Psi_{-1}\right|^2\Psi_{-2}  \notag \\
&\hspace{5mm}+\alpha_4\left|\Psi_{+1}\right|^2\Psi_{-2}+W\Psi^*_{+2}%
\Psi_{+1}\Psi_{-1}.  \label{eq:GP4}
\end{align}
Here $\hat{k}_{x,y}=-i\hat{\nabla}_{x,y}$, $m_{ex}=m_e+m_{hh}$. To make this system more compact we have omitted the terms describing exchange induced
exciton splittings given by the Hamiltonian~(\ref{eq:H4x4}). We do not
discuss here the nature and value of the interaction constants $%
\alpha_{1,2,3,4}$ and $W$. In the system of indirect excitons in coupled
GaAs/AlGaAs quantum wells, as a zeroth approximation, one can take $%
\alpha_1=\alpha_2=\alpha_3=\alpha_4$. Note also that in microcavities, where
the lower exciton-polariton mode is strongly decoupled from dark excitons,
the dark exciton states may be almost empty at low temperatures. If this is
the case, the spin dynamics of the exciton-polariton condensate is given by
the first two of the four GP equations (\ref{eq:GP1} and \ref{eq:GP2}) with $%
\alpha_{3,4}=W=0$. The remaining constants $\alpha_{1,2}$ have been widely
discussed in literature \cite{Vladimirova2010}.

The GP equations are widely used for the description of coherent propagation
of exciton-polaritons in microcavities~\cite{Wouters2010}. They allow for
the studying of interesting topology effects such as: quantum vortices~\cite%
{Whittaker2007,Liew2007,Liew2008,Marchetti2010,Pigeon2011,Lagoudakis2011};
half-quantum vortices~\cite{Rubo2007,Flayac2010,Manni2012}; bright~\cite%
{Egorov2008} and dark~\cite%
{,Yulin2008,Pigeon2011,Amo2011,Flayac2011,Ostrovskaya2012} solitons.

The polarization of light emitted by an exciton or exciton-polariton
condensate can be obtained as:
\begin{align}
\rho_c&=\frac{2S_z}{I}=\frac{\left|\Psi_{+1}\right|^2-\left|\Psi_{-1}%
\right|^2}{\left|\Psi_{+1}\right|^2+\left|\Psi_{-1}\right|^2}, \\
\rho_l&=\frac{2S_x}{I}=\frac{2\mathcal{R}e\left\{\Psi^*_{+1}\Psi_{-1}\right\}%
}{\left|\Psi_{+1}\right|^2+\left|\Psi_{-1}\right|^2}, \\
\rho_d&=\frac{2S_x}{I}=-\frac{2\mathcal{I}m\left\{\Psi^*_{+1}\Psi_{-1}\right%
\}}{\left|\Psi_{+1}\right|^2+\left|\Psi_{-1}\right|^2}.
\end{align}
%
%
%
These expressions easily follow from the definition of the spin density
matrix.

A significant limitation of the GP equations as a theoretical tool is that
they assume a coherent state of the system. If one is interested in the spin
structure of the zero temperature ground state of excitons in a
Bose-Einstein condensate~\cite{Matuszewski2012}, then this assumption is
fulfilled by definition. However, in real systems there is an incomplete
coherence that, strictly speaking, requires a description of statistical
mixtures, perhaps involving density matrices. Furthermore, Eqs.~(\ref{eq:GP1}%
-\ref{eq:GP4}) have been written assuming an infinite lifetime for the
particles (be they excitons or exciton-polaritons), which is never the case
of real non-equilibrium systems. Often pumping and radiative decay terms are
introduced into Eqs.~(\ref{eq:GP1}-\ref{eq:GP4}) phenomenologically~\cite%
{Shelykh2006,Kyriienko2012}. While in the case of a resonant coherent pump,
one can imagine that the exciton/exciton-polariton distribution inherits
coherence directly, it is less obvious how an incoherent pump can be
modelled. A phenomenological model introduced by Wouters and Carusotto~\cite%
{Wouters2007} of incoherent pumping has allowed the modelling of the first
order coherent fraction observed in many experimental configurations based
on condensation~\cite%
{Kohnle2011,Christmann2012,Tosi2012,Flayac2012,Wertz2012}. While the GP
model cannot model the phase transition during formation of a condensate
and/or superfluid, it can offer a suitable description of spin currents once
spatial coherence has formed.

Here we will consider a localized source of the four-component
indirect exciton system, as corresponds to the localized bright spot
sources generating exciton condensates~\cite{High2012}. We focus our
attention on the possible spin polarization textures of coherent
excitons propagating away from the source. We do not attempt to
describe the partially coherent state within the source, noting that
in exciton-polariton systems spin currents have been generated from
both coherent~\cite{Amo2009b} and incoherent~\cite{Kammann2012}
tightly focused spots utilizing the optical spin Hall
effect~\cite{Kavokin2005} in a similar way. 
Since the
Gross-Pitaevskii equations are only valid for coherent excitons, we
restrict the exciton wavefunction to lie outside of the source area.
The effect of the source is then characterized by the chosen
boundary condition along the edge of the source area. Given
that dark excitons have lower
energy than bright excitons [due to the exchange splitting of Eq.~(\ref%
{eq:H04x4})], it is reasonable to expect the source to
provide linearly polarized dark excitons. By fixing the values of the exciton wavefunction along the edges of the
source area, which is assumed circular, 
to such a
distribution the boundary condition
acts as an effective source for the exciton
wavefunction outside the source area.

The indirect excitons are known to have very long lifetime,
typically in the range of $10$ns to $10\mu$s \cite{Winbow2007}.
This allows them to cover distances of a few hundreds of
$\mu m$ with negligible loss~\cite{High2013}. Consequently, when we
focus on the behaviour of excitons in a small ($10\times10\mu m^2$)
area around the source, the main loss of excitons is caused by their
escape from the area of interest rather than their decay
(recombination). To model the spin currents we
thus employ an absorbing boundary condition to allow the solution of Eqs.~(%
\ref{eq:GP1}-\ref{eq:GP4}) in a finite area. 
This allows a balance between source and loss to achieve a steady state of the
non-equilibrium system (for a continuous pump), where both source and loss appear as boundary conditions.~\cite{High2013} The steady
state solution of the system is independent of the initial
condition.

\begin{figure}[h!]
\centering
\includegraphics[width=8.116cm]{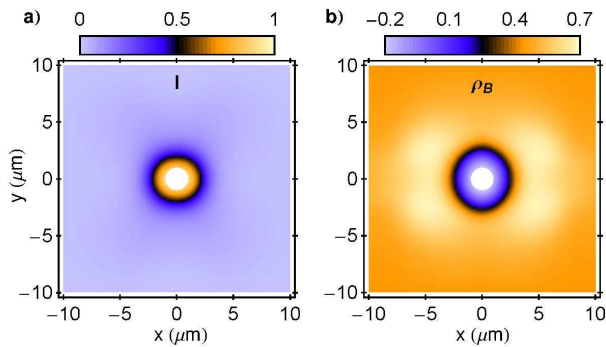}
\caption{(color online) Spatial distribution of the total exciton density
(a) and exciton brightness degree (b) [see Eq.~(\protect\ref{eq:rhob})] in
the absence of a magnetic field. The parameters were the same as those used
for the DM calculations, given in Table~\protect\ref{tab:param}, with: $B=0$%
; $W=0.2\protect\alpha$; $m_{ex}$ was taken as $0.21$ of the free electron
mass. The source area was taken circular with a radius of $1\protect\mu$m.
The images are presented with spatial averaging over $1.5\protect\mu$m. The
scattering parameter $\protect\alpha$ and intensity at the source center
were chosen such that the interaction energy, $\protect\alpha%
\left(\left|\Psi_{+1}\right|^2+\left|\Psi_{-1}\right|^2+\left|\Psi_{+2}%
\right|^2+\left|\Psi_{-2}\right|^2\right)=1\protect\mu$eV (being
comparable to the other energy scales in the system, we are in a
nonlinear regime). The absorbing boundary condition used
in calculations appears outside of the plotted range, at a radius of
$15\mu$m from the source center.}
\label{fig:GPintensityB0}
\end{figure}

\begin{figure}[h!]
\centering
\includegraphics[width=8.116cm]{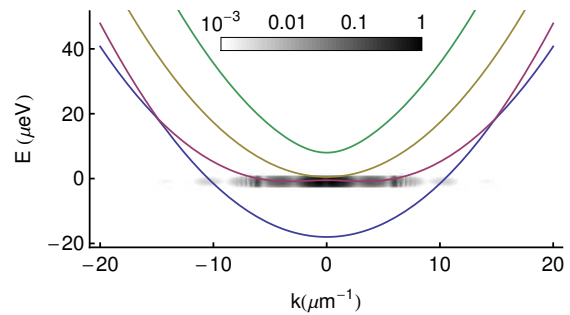}
\caption{(color online) Dispersion relation obtained from the
Gross-Pitaevskii equations, corresponding to
Figs.~\ref{fig:GPintensityB0} and \ref{fig:GPpolarizationB0}b,d and
f. The dispersion is obtained by Fourier transform of the
wavefunctions in space and time, from which the grayscale map of
intensity is obtained. The curves show the bare dispersion, obtained
from diagonalization of
$\hat{\mathcal{H}}^{tot}_{ex}+\frac{\hbar^2k^2}{2m}$.
}
\label{fig:GPDispersion}
\end{figure}

Exciton intensity and polarization distributions calculated within the GP
approach are shown in Figs.~\ref{fig:GPintensityB0} and \ref%
{fig:GPpolarizationB0}. Note that all the plotted quantities are
spatially averaged over $1.5\mu$m to account for the typical
resolution of experimental setups~\cite{High2013}. In principle
excitons can display
features on the scale of the de Broglie wavelength, $\lambda=2\pi/|\mathbf{%
k_{ex}}|$. Such features are on the sub-micron scale and are far
beyond experimental resolution. We note also that while in the DM
approach we could consider excitons having a fixed radial velocity,
in the GP approach we necessarily cover the whole range of
wavevectors and propagation velocities (the dispersion obtained from
the GP approach is shown in Fig.~\ref{fig:GPDispersion}). Also,
exciton-exciton interactions modify the exciton dispersion. For all
these reasons, the DM and GP approaches cannot give identical
results, in principle. The DM approach is suitable for the
description of both coherent and partly coherent exciton gases,
while the GP approach better catches the dispersive propagation
features and accounts for nonlinear effects. Both approaches are
complementary, and it is instructive to compare the results,
obtained within these two models.

The exciton density decreases as excitons travel away from the source (Fig.~%
\ref{fig:GPintensityB0}a). This is not due to exciton recombination, which
is expected to be very slow, but more simply due to the spreading out of
excitons in all directions. The intensity spread need not be perfectly
circularly symmetric due to the presence of the spin-orbit (Dresselhaus
terms), which can introduce a directional dependence of the exciton
velocity. Fig.~\ref{fig:GPintensityB0}b shows the exciton brightness degree,
defined as:
\begin{equation}
\rho_b=\frac{\left|\Psi_{+1}\right|^2+\left|\Psi_{-1}\right|^2-\left|%
\Psi_{+2}\right|^2-\left|\Psi_{-2}\right|^2}{\left|\Psi_{+1}\right|^2+\left|%
\Psi_{-1}\right|^2+\left|\Psi_{+2}\right|^2+\left|\Psi_{-2}\right|^2}.
\label{eq:rhob}
\end{equation}
This quantity represents the degree to which the bright exciton density
exceeds the dark exciton density. There is a conversion of dark to bright
excitons as they spread out from the source, which can be expected from the
presence of Dresselhaus coupling terms.

\begin{figure}[h]
\centering
\includegraphics[width=8.116cm]{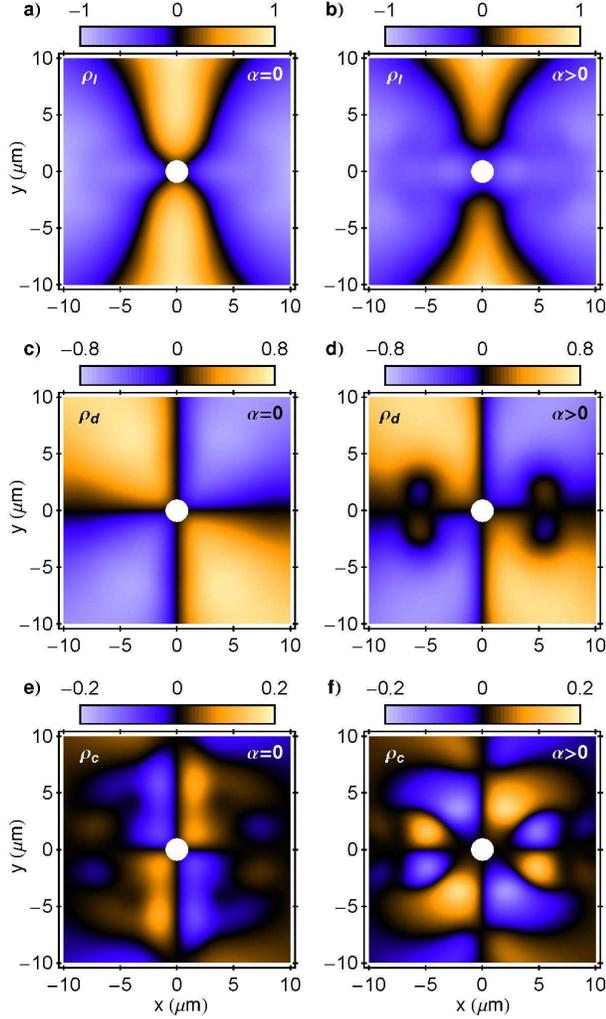}
\caption{(color online) Spatial distribution of the polarization state of
excitons in the absence of a magnetic field: horizontal-linear polarization
degree (a, b), diagonal polarization degree (c, d) and circular polarization
degree (e, f) for bright excitons, which corresponds to the near field
emission pattern of light. The left-hand plots show results in the absence
of nonlinear interactions ($\protect\alpha =0$; $W=0$), while the right-hand
plots show the case of a moderate nonlinearity. The parameters were the same
as in Fig.~\protect\ref{fig:GPintensityB0}}
\label{fig:GPpolarizationB0}
\end{figure}

The polarization distribution is shown in Fig.~\ref{fig:GPpolarizationB0}
and can be significantly influenced by non-linearity in the system. The
left-hand plots show the results for negligible nonlinearity ($\alpha=0$; $%
W=0$), which is equivalent to a weak pump intensity. Here the polarization
distributions are qualitatively similar to those calculated in the density
matrix formalism. In analogy to the (intrinsic) spin Hall effect~\cite%
{Sinova2004,Mishchenko2004} and the optical spin Hall effect~\cite%
{Kavokin2005,Leyder2007} the presence of the spin-orbit coupling terms
introduces a directional dependence of the polarization. The patterns of the
polarization degrees divide into quadrants. Some quantitative differences
with the DM calculations appear due to the presence of different wavevectors.

The right-hand plots show the case of a moderate nonlinearity, with
interaction strength comparable to the other energy scales of the
system. The most drastic effect is on the circular polarization
degree, which becomes higher and each quadrant of circular
polarization divides further giving an eight-lobed pattern to the
polarization degree. The interaction terms that we have introduced
are all spin conserving and it can be noted that we have considered
the spin isotropic case. Even the $W$ nonlinear interaction term,
which allows the inter-conversion of bright and dark exciton pairs
does not appear to directly change the spin polarization, conserving
both circular and linear polarizations upon scattering. Still, the
nonlinear interaction terms can have a drastic effect on the
polarization structure. This is because they are able to shift
(renormalize) the dispersion branches in the system. Given that the
potential energy of excitons is fixed by their interaction energy at
the source and that this energy is converted into kinetic energy at
distances away from the source, any shifts in the dispersion
branches can change the wavevector of propagating excitons. Even if
the nonlinear induced shifts of the dispersion branches were not
polarization dependent, a change in the wavevector of an exciton can
allow it to experience a different effect from the k-dependent
spin-orbit coupling terms. In this way, richer structures can appear
in the nonlinear regime. Note, that the build-up of circular
polarisation clearly seen in Fig.~\ref{fig:GPpolarizationB0} (e,f)
would not yield 100\% circularly polarised excitons, as it might be
expected in the ideal case of the optical spin Hall effect for
exciton-polaritons \cite{Kavokin2005}. In our system, the precession
of electron and hole spins has different frequencies, and the
interplay between dark-to-bright and linear-to-circular polarisation
conversion prevents formation of a purely circularly polarised
state.

\section{Exciton spin currents}

Consider an exciton state characterized by a wavevector $\mathbf{k}_{ex}$
and described by the DM $\hat{\rho}$. Let us recall that the elements of
this matrix $\rho _{11}$, $\rho _{22}$, $\rho _{33}$, $\rho _{44} $ are the
densities of +1,-1,+2 and -2 spin polarized excitons, respectively. The
current of each of these densities is given by a product of the exciton
speed and the corresponding density:
\begin{equation}
\mathbf{j}_{a}=\frac{\hbar \mathbf{k}_{ex}}{m_{ex}}\rho _{jj}
\label{eq:jalpha}
\end{equation}%
with $j=1,2,3,4$ for $a=+1,-1,+2,-2$, respectively. Experimentally, one can
measure the magnetization current associated with the exciton density
current. The magnetization carried by propagating excitons can be found as:
\begin{equation}
M_{z}=-\frac{\mu _{B}}{2\hbar }\left[ \left( g_{h}-g_{e}\right) \left( \rho
_{11}-\rho _{22}\right) +\left( g_{h}+g_{e}\right) \left( \rho _{33}-\rho
_{44}\right) \right]
\end{equation}%
This expression is obtained having in mind that an electron with a spin
projection on the z-axis of $\pm 1/2$ contributes to the magnetization
projection on the z-axis $\mp \frac{\mu _{B}}{2}g_{e}$, and a heavy hole
with the spin projection of $\pm 3/2$ contributes to the magnetization $\mp
\frac{\mu _{B}}{2}g_{h}$. Hence, the magnetization (spin) current produced
by the excitons having a wave-vector $\mathbf{k}_{ex}$ will be given by:
\begin{align}
\mathbf{j}_{M}(\mathbf{k}_{ex})& =-\frac{\mu _{B}\mathbf{k}_{ex}}{2m_{ex}}%
\left[ \left( g_{h}-g_{e}\right) \left( \rho _{11}(\mathbf{k}_{ex})-\rho
_{22}(\mathbf{k}_{ex})\right) \right.  \notag \\
& \hspace{10mm}+\left. \left( g_{h}+g_{e}\right) \left( \rho _{33}(\mathbf{k}%
_{ex})-\rho _{44}(\mathbf{k}_{ex})\right) \right]  \label{eq:jm}
\end{align}%
The total magnetization current in the exciton gas can be obtained by
integration over all wave-vectors:
\begin{equation}
\mathbf{j}_{M}^{tot}=-\frac{A}{\left( 2\pi \right) ^{2}}\int {\mathbf{j}_{M}(%
\mathbf{k}_{ex})d\mathbf{k}_{ex}}  \label{eq:jMtot}
\end{equation}%
Here $A$ is the area of the sample. This current may be detected, for
example, by spatially resolved Kerr rotation spectroscopy.

\begin{figure}[h]
\centering
\includegraphics[width=8.116cm]{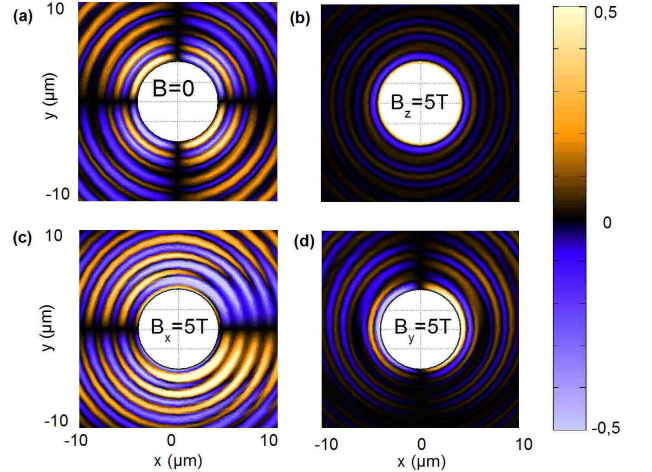}
\caption{(color online) Spatial distribution of the exciton spin current
calculated in arbitrary units using the parameters summarized in Table~%
\protect\ref{tab:param} at $\mathbf{k}_{ex}=15.3\protect\mu$m$^{-1}$ at $B=0$
(a) and $B=5$T (b)-(d) along the z, x and y-axis, respectively. The source
area was taken circular with a radius of $4\protect\mu$m.}
\label{Fig_jm}
\end{figure}

Figure \ref{Fig_jm} shows the spin current density $\mathbf{j}_{M}/\left(
2\pi r\right) $ calculated for the system of indirect excitons which we
considered above in the absence of external magnetic field (a) and in the
presence of a magnetic field of 5T oriented normally to the plane of the
structure (b) and along $x$- and $y$-axes (c,d). In all cases the current
intensity decreases as one moves away from the excitation spot, as the
exciton density decreases inversely with the radius $r$. One can see that
the total spin (magnetization) of propagating excitons experiences
oscillations and has a strong angular dependence. The spin currents are
suppressed by an in-plane magnetic field, which is not surprising: $\mathbf{j%
}_{M}$ describes propagation of the normal-to plane spin component, which is
strongly reduced by an in-plane magnetic field. The images show the total
spin carried by both bright and dark excitons. They do not directly
correspond to the polarized photoluminescence map for two reasons: first,
dark excitons do not contribute to the photoluminescence; second, the
polarization degree of photoluminescence does not experience the 1/$r$ decay
characteristic of the total spin density. On the other hand, the images
presented in Figure \ref{Fig_jm} do correspond to the signal of spatially
resolved Kerr or Faraday rotation, which is sensitive to the normal-to-plane
magnetization.

\section{Spin currents in exciton condensates}

The approach formulated above can be extended to the description of spin
currents in coherent exciton (or exciton-polariton) condensates accounting
for particle-particle interactions. In this case we need to replace the
momentum $\hbar \mathbf{k}_{ex}$ by a momentum operator $\hat{p}=-i\hbar
\hat{\nabla}$ and the diagonal components of the DM $\rho _{11}$, $\rho
_{22} $, $\rho _{33}$, $\rho _{44}$ by the exciton densities $|\psi
_{+1}|^{2},|\psi _{-1}|^{2},|\psi _{+2}|^{2},|\psi _{-2}|^{2}$,
respectively, in the expressions~\ref{eq:jalpha} and \ref{eq:jMtot}. In this
case the density currents become:
\begin{equation}
\mathbf{j}_{\alpha }=-i\frac{\hbar }{m_{ex}}\Psi _{\alpha }^{\ast }\nabla
\Psi _{\alpha },  \label{eq:jalphapsi}
\end{equation}%
and the total magnetization current can be expressed as:
\begin{align}
\mathbf{j}_{M}^{tot}& =\frac{i\mu _{B}}{2m_{ex}}\left[ \left(
g_{h}-g_{e}\right) \left( \Psi _{+1}^{\ast }\nabla \Psi _{+1}-\Psi
_{-1}^{\ast }\nabla \Psi _{-1}\right) \right.  \notag \\
& \hspace{15mm}+\left. \left( g_{h}+g_{e}\right) \left( \Psi _{+2}^{\ast
}\nabla \Psi _{+2}-\Psi _{-2}^{\ast }\nabla \Psi _{-2}\right) \right] \\
& =-\frac{\mu _{B}}{2\hbar }\left[ \left( g_{h}-g_{e}\right) \left( \mathbf{j%
}_{+1}-\mathbf{j}_{-1}\right) \right.  \notag \\
& \hspace{15mm}+\left. \left( g_{h}+g_{e}\right) \left( \mathbf{j}_{+2}-%
\mathbf{j}_{-2}\right) \right]  \label{eq:jMtotpsi}
\end{align}

\begin{figure}[h!]
\centering
\includegraphics[width=8.116cm]{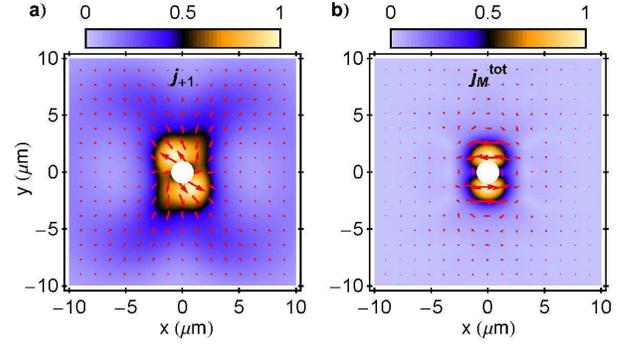}
\caption{(color online) Spatial structure of the spin density current $%
\mathbf{j}_{+1}$ (a) and the total magnetization current $\mathbf{j}_M^{tot}$
(b). The arrows show the directional dependence of the vector fields in
space, while the colour code illustrates the intensity. The parameters were
the same as in Fig.~\protect\ref{fig:GPintensityB0}. Arbitrary units are
used for both the spin density and total magnetization current.}
\label{fig:GPcurrentsB0}
\end{figure}

The distribution of the spin density current, $\mathbf{j}_{+1}$, is shown in
Fig.~\ref{fig:GPcurrentsB0}~(a). The current density propagates outward from
the source in all directions, decreasing in intensity. The apparent rotation
of the current density is a nonlinear effect coming from the interactions in
the system. The other spin density currents, $\mathbf{j}_{-1}$, $\mathbf{j}%
_{+2}$ and $\mathbf{j}_{+2}$, display a similar behaviour.

The magnetization current is shown in Fig.~\ref{fig:GPcurrentsB0}~(b). The
current is stronger closer to the source, where the intensities are
stronger. The magnetization current is predicted to rotate around the source.

One can also introduce the spin conductivity tensor linking the components
of the density current (\ref{eq:jalphapsi}) with the gradient of potential
acting upon each of the exciton spin components:
\begin{equation}
j_{\alpha,l}=\sigma^{l,m}_{\alpha,\beta}\nabla U_{\beta,m},
\end{equation}
where $l=x,y$ and $m=x,y$ indicate the in-plane projections of the current
and potential gradient, respectively. One can see that $\sigma^{l,m}_{%
\alpha,\beta}$ is a 64-component tensor in the general 2D case. The origin
of the potential gradient $\nabla U_{\beta,m}$ needs to be discussed
separately. 
$\nabla U_{\beta,m}$ can originate from the gradient of the quantum well
width, gradient of the barrier height, or it can be induced by excitons
themselves due to e.g. dipole-dipole repulsion. Indirect excitons have
built-in dipole moments, the laterally modulated external electric field in
the z-direction can create an in-plane potential landscape and, in turn,$%
\nabla U$ for them. This was used in studies of transport of indirect
excitons in various electrostatic potential landscapes including potential
energy gradients \cite{Hagn1995, Gartner2006, Leonard2012}, circuit devices
\cite{High2007, High2008, Grosso2009}, traps \cite{High2009} lattices \cite%
{Remeika2009, Remeika2012}, moving lattices-conveyers \cite{Winbow2011}, and
narrow channels \cite{Grosso2009, Vogele2011, Cohen2011}.

\section{Polarization currents}

Spatially resolved measurements of the polarization degrees $\rho_c$, $%
\rho_l $ and $\rho_d$ of light emitted by excitons give access to the
exciton polarization currents. In terms of the DM formalism, they can be
defined as products of the exciton speed and the corresponding polarization
degree:
\begin{align}
\mathbf{j}_c\left(\mathbf{k}_{ex}\right)&=\frac{\hbar\mathbf{k}_{ex}}{m_{ex}}%
\rho_c=\frac{\hbar\mathbf{k}_{ex}}{m_{ex}}\frac{\rho_{11}-\rho_{22}}{%
\rho_{11}+\rho_{22}}  \label{eq:jc} \\
\mathbf{j}_l\left(\mathbf{k}_{ex}\right)&=\frac{\hbar\mathbf{k}_{ex}}{m_{ex}}%
\rho_l=\frac{\hbar\mathbf{k}_{ex}}{m_{ex}}\frac{\rho_{12}+\rho_{21}}{%
\rho_{11}+\rho_{22}}  \label{eq:jl} \\
\mathbf{j}_d\left(\mathbf{k}_{ex}\right)&=\frac{\hbar\mathbf{k}_{ex}}{m_{ex}}%
\rho_d=\frac{\hbar\mathbf{k}_{ex}}{m_{ex}}\frac{\rho_{12}-\rho_{21}}{%
\rho_{11}+\rho_{22}}  \label{eq:jd}
\end{align}

The total polarization currents can be obtained integrating the expressions (%
\ref{eq:jc}-\ref{eq:jd}) over reciprocal space:
\begin{equation}
\mathbf{j}_{tot}^{c,l,d}=-\frac{A}{\left(2\pi\right)^2}\int{d\mathbf{k}_{ex}%
\mathbf{j}_{c,l,d}\left(\mathbf{k}_{ex}\right)}
\end{equation}

The polarization currents in an exciton condensate can be found from the GP
equations (\ref{eq:GP1}-\ref{eq:GP4}) as:
\begin{align}
\mathbf{j}_c&=-\frac{i\hbar}{m_{ex}}\frac{\left(\Psi_{+1}^*\nabla\Psi_{+1}-%
\Psi^*_{-1}\nabla\Psi_{-1}\right)}{|\Psi_{+1}|^2+|\Psi_{-1}|^2}
\label{eq:jcpsi} \\
\mathbf{j}_l&=-\frac{i\hbar}{m_{ex}}\frac{\left(\Psi_{+1}^*\nabla\Psi_{-1}+%
\Psi^*_{-1}\nabla\Psi_{+1}\right)}{|\Psi_{+1}|^2+|\Psi_{-1}|^2}
\label{eq:jlpsi} \\
\mathbf{j}_d&=-\frac{\hbar}{m_{ex}}\frac{\left(\Psi_{+1}^*\nabla\Psi_{-1}-%
\Psi^*_{-1}\nabla\Psi_{+1}\right)}{|\Psi_{+1}|^2+|\Psi_{-1}|^2}
\label{eq:jdpsi}
\end{align}

\begin{figure}[h!]
\centering
\includegraphics[width=8.116cm]{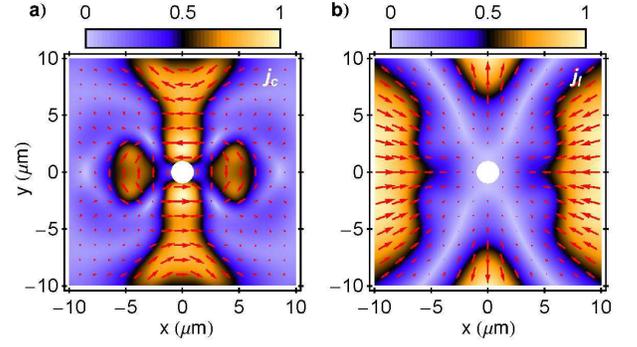}
\caption{(color online) Spatial structure of the polarization currents $%
\mathbf{j}_c$ (a) and $\mathbf{j}_l$ (b). The arrows show the directional
dependence of the vector fields in space, while the colour code illustrates
the intensity. The parameters were the same as in Fig.~\protect\ref%
{fig:GPintensityB0}. Arbitrary units are used for the polarization currents.}
\label{fig:GPpolcurrentsB0}
\end{figure}

The distributions of the polarization currents calculated within GP approach
are shown in Fig.~\ref{fig:GPpolcurrentsB0}. A striking nonuniform structure
appears due to the presence of the spin-orbit coupling terms. Close to the
source spot, there is a strong circularly polarized current that rotates
around the source. This can be attributed to the rotating circular
polarization degree already observed in Fig.~\ref{fig:GPpolarizationB0}.
Away from the source, the circular polarization current decays, which can be
expected due to the decay of the spin density current observed in Fig.~\ref%
{fig:GPcurrentsB0}. Along the vertical axis ($x=0$), a strong circular
polarization current remains due to the particularly fast change of the
circular polarization degree in this region. The linearly polarized current
can be stronger further away from the source than at closer distances. This
is attributed to an increasing linear polarization degree further from the
source. In addition, one can recall that while spin density currents are
generally weaker further from the source, there is some compensation due to
the conversion between dark and bright excitons (as shown in Fig.~\ref%
{fig:GPintensityB0}b, the bright exciton fraction increases further from the
source).

\section{Conclusions}

Bosonic spin transport is a young and promising area of solid-state physics.
The theories of mesoscopic transport of charge carriers and quantum
transport are among the most interesting chapters of modern physics.
Substitution of fermions by bosons and of a scalar electric charge by a spin
vector cannot be formally done in these theories. Basically, all mesoscopic
and quantum transport effects need to be reconsidered if we speak about
electrically neutral bosonic spin carriers like excitons or
exciton-polaritons. 
This is why the area of \textquotedblleft spin-optronics\textquotedblright\
essentially remains \textit{terra incognita}. Experimentally, direct
measurements of transport of indirect excitons and exciton-polaritons in
time-resolved imaging experiments have become possible in recent years. %
%
In this work, we have demonstrated that exciton polarization currents are
inseparably connected with electron and hole spin currents. The intensity
and direction of exciton polarization currents and electron and hole spin
currents is governed by an interplay of spin-orbit effects, Zeeman effects
and exciton exchange effects. In the non-linear regime, the pattern of spin
currents may also be affected by spin-dependent exciton-exciton interactions.

We have developed two complementary approaches to the description of exciton
spin currents and textures. The DM formalism allows for description of the
spin transport effects in both classical exciton gases and condensates of
non-interacting excitons, while the GP equations describe propagation of
exciton condensates. We predict non-trivial topologies of interacting
exciton spin in condensates, and suggest tools of their control, such as
external magnetic and electric fields, and source intensity. We have
demonstrated, that ballistic propagation of excitons may result in a build
up of polarization patterns, which may be observed in near-field
photoluminescence spectra. 

This work has been supported by the EU FP7 ITN INDEX, EU FP7 PodiTrodi, the
E.P.S.R.C., the EU FP7 Marie Curie EPOQUES, ANR-2011-NANO-004-06, and DOE.
J. R. L. acknowledges Chateaubriand Fellowship from the Embassy of France in
the United States. The authors are deeply grateful to T. Ostatnicky, Y.G.
Rubo, M.M. Glazov, I.A. Shelykh, and A. Bramati for many useful discussions
on the peculiarities of bosonic spin transport.

\end{document}